\documentclass{aa}  
\usepackage{graphicx}
\usepackage{txfonts}
\usepackage{natbib}
\bibpunct{(}{)}{;}{a}{}{,} 

   \newcommand{\beam}{$\theta_{\mbox{\scriptsize maj}}\times\theta_{\mbox{\scriptsize min}}$}

\begin{document}

   \title{Forming localized dust concentrations in a dust ring: DM Tau case study}

   \subtitle{The asymmetric 7~mm dust continuum of the DM~Tau disk}

   \author{Hauyu Baobab Liu
          \inst{1,2}
          \and
          Takayuki Muto\inst{3,4,5}
          \and
          Mihoko Konishi\inst{6}
          \and
          Chia-Ying Chung\inst{7,1}
          \and
          Jun Hashimoto\inst{8,9,10}
          \and
          Kiyoaki Doi\inst{11,12}
          \and
          Ruobing Dong\inst{13}
          \and
          Tomoyuki Kudo\inst{14}
          \and
          Yasuhiro Hasegawa\inst{15}
          \and
          Yuka Terada\inst{7,1}
          \and
          Akimasa Kataoka\inst{17}
          }

   \institute{
   Department of Physics, National Sun Yat-Sen University, No. 70, Lien-Hai Road, Kaohsiung City 80424, Taiwan, R.O.C.\\
              \email{baobabyoo@gmail.com}
         \and
    Center of Astronomy and Gravitation, National Taiwan Normal University, Taipei 116, Taiwan
         \and
    Division of Liberal Arts, Kogakuin University, 1-24-2, Nishi-Shinjuku, Shinjuku-ku, Tokyo 163-8677, Japan
        \and
    Leiden Observatory, Leiden University, P.O. Box 9513, NL-2300 RA Leiden, The Netherlands
        \and
    Department of Earth and Planetary Sciences, Tokyo Institute of Technology, 2-12-1 Oh-okayama, Meguro-ku, Tokyo 152-8551, Japan
        \and
    Faculty of Science and Technology, Oita University, 700 Dannoharu, Oita, 870-1192, Japan
        \and
    Academia Sinica Institute of Astronomy and Astrophysics (ASIAA), No. 1, Section 4, Roosevelt Road, Taipei 10617, Taiwan
        \and
    Astrobiology Center, National Institutes of Natural Sciences, 2-21-1 Osawa, Mitaka, Tokyo 181-8588, Japan
        \and
    Subaru Telescope, National Astronomical Observatory of Japan, Mitaka, Tokyo 181-8588, Japan
        \and
    Department of Astronomy, School of Science, Graduate University for Advanced Studies (SOKENDAI), Mitaka, Tokyo 181-8588, Japan
        \and
    Department of Astronomical Science, School of Physical Sciences, Graduate University for Advanced Studies (SOKENDAI), 2-21-1 Osawa, Mitaka, Tokyo 181-8588, Japan
        \and
    National Astronomical Observatory of Japan, 2-21-1 Osawa, Mitaka, Tokyo 181-8588, Japan
        \and
    Department of Physics \& Astronomy, University of Victoria, Victoria, BC, V8P 5C2, Canada
        \and
    SubaruTelescope, National Astronomical Observatory of Japan, 650 North A$'$ohoku Place, Hilo, HI 96720  U.S.A
        \and
    Jet Propulsion Laboratory, California Institute of Technology, Pasadena, CA 91109, USA
        \and
    Department of Astrophysics, National Taiwan University, Taipei 10617, Taiwan, R.O.C.
        \and
    National Astronomical Observatory of Japan, 2-21-1 Osawa, Mitaka, Tokyo 181-8588, Japan
             }

   \date{Received September 15, 1996; accepted March 16, 1997}

 
  \abstract
   {Previous high-angular-resolution 225 GHz ($\sim$1.3 mm) continuum observations of the transitional disk DM~Tau  have resolved an outer ring at 20-120 au radii that is weakly azimuthally asymmetric. }
   {We aim to examine dust growth and filtration in the outer ring of DM Tau.}
   {We performed $\sim$0\farcs06 ($\sim$8.7 au) resolution Karl G. Jansky Very Large Array (JVLA) 40--48 GHz ($\sim 7$ mm; Q band) continuum observations, along with complementary observations at lower frequencies. 
In addition, we analyzed the archival JVLA observations undertaken since 2010.}
   {Intriguingly, the Q band image resolved the azimuthally highly asymmetric, knotty dust emission sources close to the inner edge of the outer ring. Fitting the 8-700 GHz spectral energy distribution (SED) with two dust components indicates that the maximum grain size ($a_{\mbox{\scriptsize max}}$) in these knotty dust emission sources is likely $\gtrsim$300 $\mu$m, whereas it is $\lesssim$50 $\mu$m in the rest of the ring. These results may be explained by a trapping of inwardly migrating "grown" dust close to the ring inner edge.
The exact mechanism for developing the azimuthal asymmetry has not yet been identified, which may be due to planet-disk interaction that might also be responsible for the creation of the dust cavity and pressure bump. Otherwise, it may be due to the fluid instabilities and vortex formation as a result of shear motions. 
Finally, we remark that the asymmetries in DM~Tau are difficult to diagnose from the $\gtrsim$225 GHz observations, owing to a high optical depth at the ring. In other words, the apparent symmetric or asymmetric morphology of the transitional disks may be related to the optical depths of those disks at the observing frequency. 
}
   {}

   \keywords{Protoplanetary disks --
                Planets and satellites: formation --
                (ISM:) dust, extinction --
                Radio continuum: ISM
               }

   \maketitle
%

\section{Introduction} \label{sec:intro}

The rich structures in protoplanetary disks resolved in recent (sub)millimeter interferometric and near infrared imaging observations have sparked a surge in theoretical follow-up studies (for a review, see \citealt{Andrews2020ARA&A..58..483A}). 
In particular, intense discussions on the origin and subsequent evolution of the concentric gaps and rings seen in a large number of protoplanetary disks has been ongoing (e.g., \citealt{Andrews2011ApJ...732...42A,Hashimoto2012ApJ...758L..19H,ALMA2015ApJ...808L...3A,Andrews_2018}).
These structures may have been created as a result of ice lines (e.g., \citealt{Zhang2015ApJ...806L...7Z,Okuzumi2016ApJ...821...82O}), planet-disk interaction (\citealt{Zhang2015ApJ...806L...7Z,Okuzumi2016ApJ...821...82O}), secular gravitational instability (e.g., \citealt{Takahashi2014ApJ...794...55T}), or other mechanisms.

\begin{figure*}
    \begin{tabular}{ p{8.5cm} p{8.5cm} }
    \includegraphics[height=10cm]{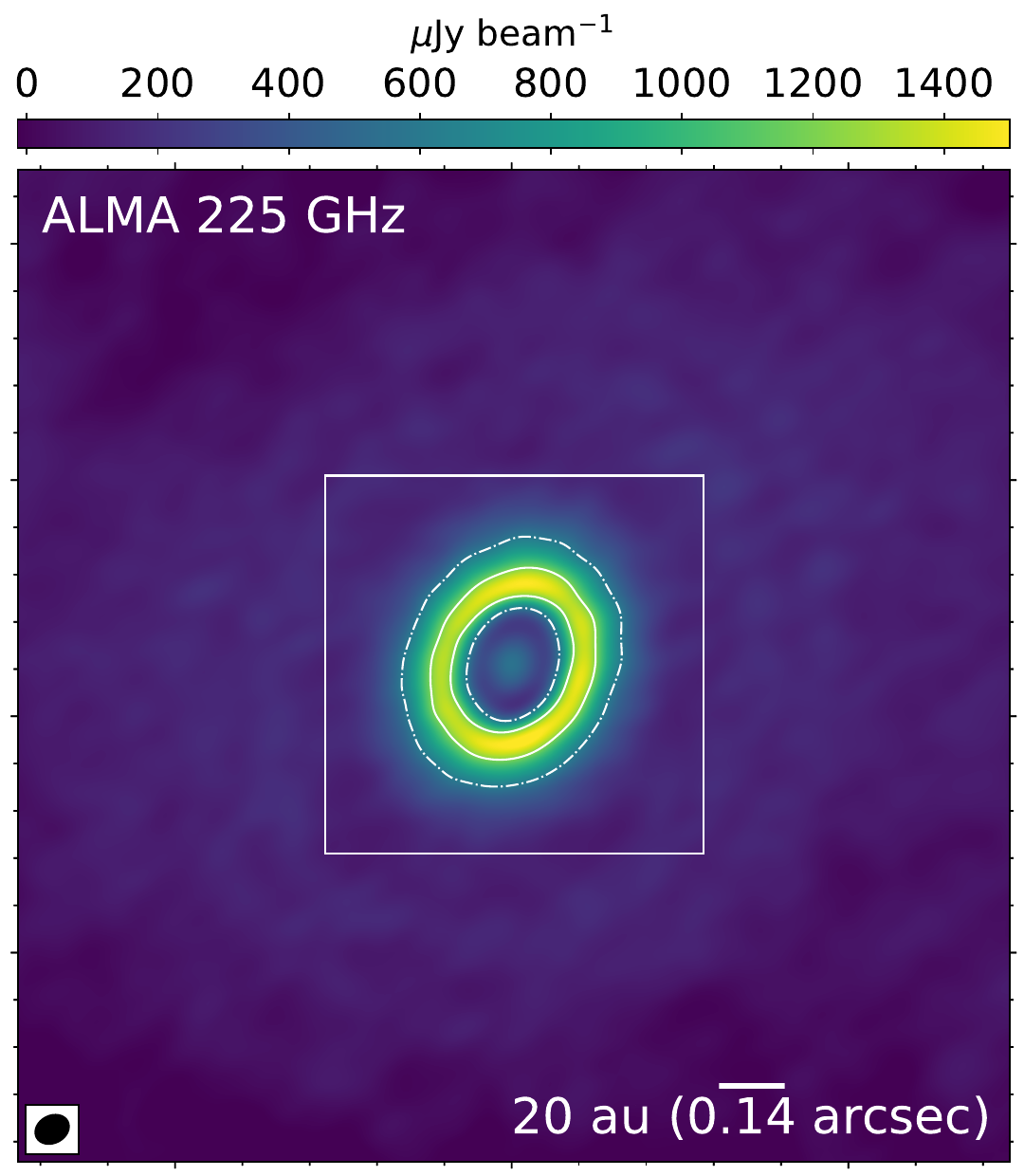} &
    \includegraphics[height=10cm]{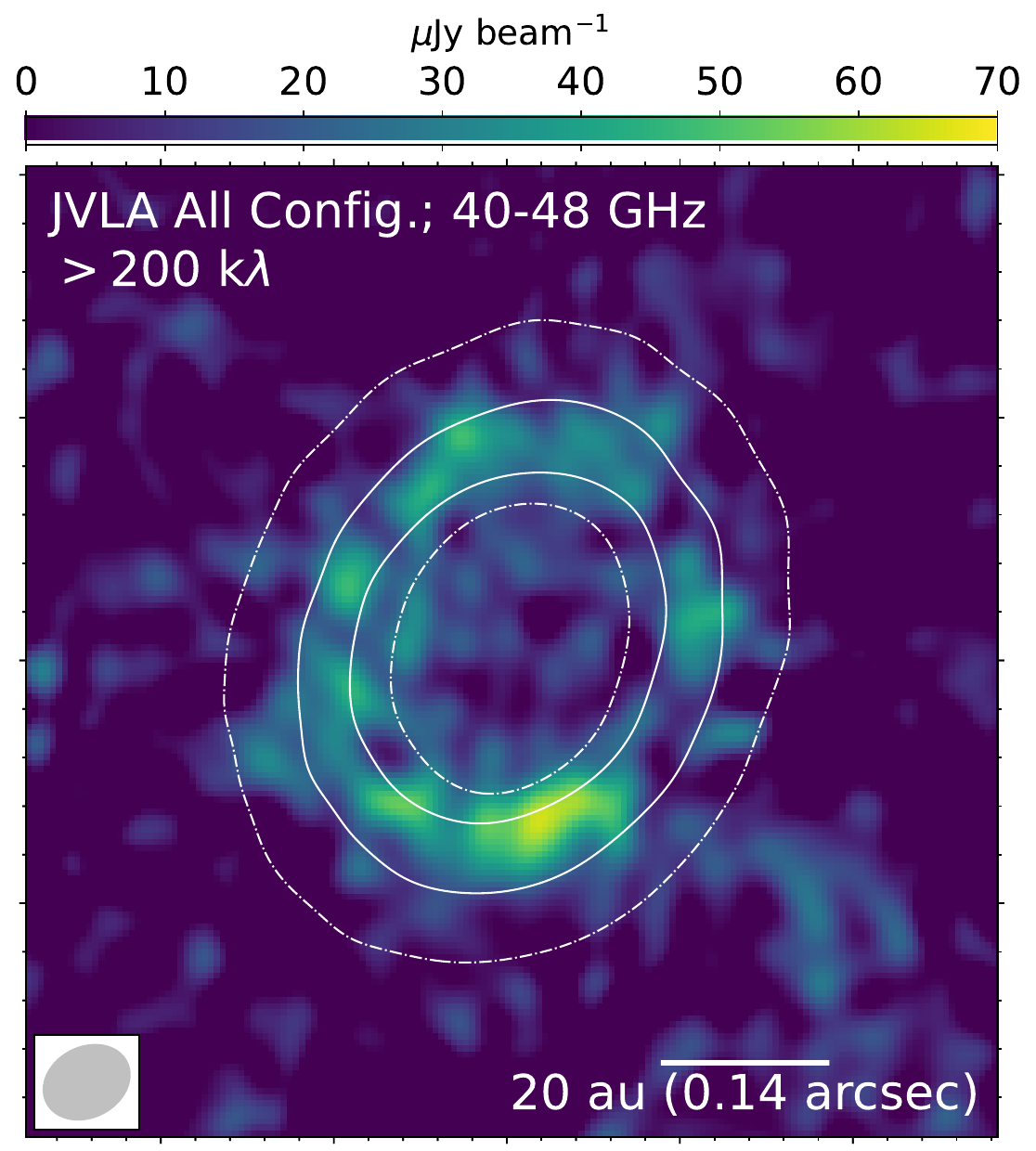} \\
    \end{tabular}
    \caption{
    ALMA and JVLA images on DM~Tau.
    {\it Left:} Previously published ALMA 225 GHz ($\sim$1.3 mm) continuum image (color and contours; \citealt{Kudo2018ApJ...868L...5K,Hashimoto2021ApJ...911....5H})  smoothed to the synthesized beam of the JVLA image in the right panel. Dash-dotted and solid contours are 600 and 1200 $\mu$Jy\,beam$^{-1}$, respectively.
    {\it Right:}  JVLA 40--48 GHz ($\sim$7 mm) continuum image (color; \beam$=$0\farcs074$\times$0\farcs058, P.A.$=-$63$^{\circ}$; RMS$=$11 $\mu$Jy\,beam$^{-1}$) in the region enclosed by the white box in the left panel. This image was created by jointly imaging all Q band observations listed in Table \ref{tab:obs}, limiting the {\it uv}-distance to $>$200 k$\lambda$.
    The images made with other combination of array configurations are provided in Figure \ref{fig:moreimage}.
    Contours are identical to those plotted in the left panel. 
    The synthesized beams of the ALMA (black) and JVLA images (gray) are plotted in the bottom-left.
    }
    \vspace{0.37cm}
    \label{fig:image}
\end{figure*}

DM Tau is a 0.53 M$_{\odot}$ star ($d\sim$145 pc; \citealt{Gaia_2023}) with an effective temperature of 3705 K  (\citealt{Kudo2018ApJ...868L...5K}).
 Analyses of its near infrared spectral energy distribution (NIR SED) have indicated that DM~Tau hosts a transitional disk that have a large inner cavity (\citealt{Bergin2004ApJ...614L.133B,Calvet2005ApJ...630L.185C}).
The 8--12 GHz and 12--18 GHz images (\citealt{Terada2023ApJ...953..147T}) as well as the 8--10 GHz non-detection (\citealt{Zapata2017ApJ...834..138Z}) have indicated that there are time varying free-free emission sources (e.g., ionized disk) within the dust cavity.
The (sub)millimeter interferometric imaging observations of dust continuum emission using the Submillimeter Array (SMA) and the Atacama Large Millimeter/submillimeter Array (ALMA) have resolved the radially extended outer dusty ring at $\sim$20--120 au radii (\citealt{Andrews2011ApJ...732...42A,Kudo2018ApJ...868L...5K}).
The ALMA $\sim$225 GHz observations additionally resolved an inner ring at $\sim$4 au radii, which appears to be azimuthally asymmetric \citep{Kudo2018ApJ...868L...5K}. 
The higher angular resolution ALMA $\sim$225 GHz continuum image further reveals a weak azimuthal asymmetry in the outer ring (\citealt{Hashimoto2021ApJ...911....5H}).
The dust mass and dust growth in the outer ring, which are tightly related to the formation of comets and ice giants, have not yet been constrained by optically thin observations.

In this work, we present the high-resolution National Radio Astronomy Observatory (NRAO) Karl G. Jansky Very Large Array (JVLA) observations of DM~Tau, which is advantageous for probing $\gtrsim$1 mm sized dust.
Besides examining the JVLA images, we based our study on fittings of the 8--700 GHz SED measured by JVLA, SMA, and ALMA to quantitatively constrain the dust column densities ($
\Sigma_{\mbox{\scriptsize dust}}$) and dust maximum grain sizes ($a_{\mbox{\scriptsize max}}$).
We introduce the data we utilized and the data calibration procedures in
Section \ref{sec:observations}.
The results are presented in Section \ref{sec:results}.
In Section \ref{sec:discussion}, we discuss the origin of the emission using SED models, and address the potential physical implication.
A brief conclusion is given in Section \ref{sec:conc}.
Appendix \ref{appendix:image} provides a comparison of the JVLA 40--48 GHz images ($\sim$7 mm) carried out via a range of approaches.
Appendix \ref{appendix:asym} details a model-independent approach to diagnose azimuthal asymmetry.
We assess the degree of free-free contamination in the JVLA 40--48 GHz images in Appendix \ref{appendix:freefree}.
Appendix \ref{appendix:mcmc} provides the details of how we modeled the observed SED using the Markov chain Monte Carlo (MCMC) method.


\section{Data and reduction}\label{sec:observations}
\subsection{Data}\label{sec:data}

We carried out the JVLA observations towards DM~Tau at Q (40--48 GHz), Ku (12--18 GHz), and X (8--12 GHz) bands in 2019.
The Ku and X bands observations which mainly traced free-free emission have been reported elsewhere \citep{Terada2023ApJ...953..147T}.
In addition, we retrieved and analyzed the archival JVLA data which were taken since August 22, 2010. 
Most of those data were taken from the Disk@EVLA project (project code: AC982; PI: C.\ Chandler), while some centimeter band data were taken from the Gould's Belt Very Large Array Survey project (project code: BL175, PI: L.~R.\ Loinard; \citealt{Dzib2015ApJ...801...91D}) and project 13B-364 (PI: L.~F.\ Rodriguez; \citealt{Zapata2017ApJ...834..138Z}). 
These JVLA observations took full RR, RL, LR, and LL correlator products. 
Table~\ref{tab:obs} summarizes the observational details and our data quality assessment. 

We performed a $\sim$3$''$ angular resolution SMA 200--400 GHz survey towards 47 Class II objects in the Taurus-Auriga region in 2021 that included DM~Tau as one of the target sources. 
The full survey will be reported in an upcoming paper (Chung et al. submitted).
The present work utilizes the 200--400 GHz SED provided by this SMA survey, which was carefully calibrated. 

We retrieved the archival ALMA Band 3 ($\sim$95--111 GHz), Band 4 ($\sim$144--159 GHz), and Band 9 ($\sim$659--676 GHz) observations that the maximum recoverable angular scales (MAS) are larger than 3$''$ for the purpose of deriving the (sub)millimeter SEDs.
We visually inspected the spectra and only utilized the passbands that detected DM~Tau and did not present significant spectral line emission.
Table \ref{tab:ALMAarchive} summarizes the ALMA measurements we included. 

\begin{figure}
    \hspace{-0.75cm}
    \begin{tabular}{c}
    \includegraphics[width=10.5cm]{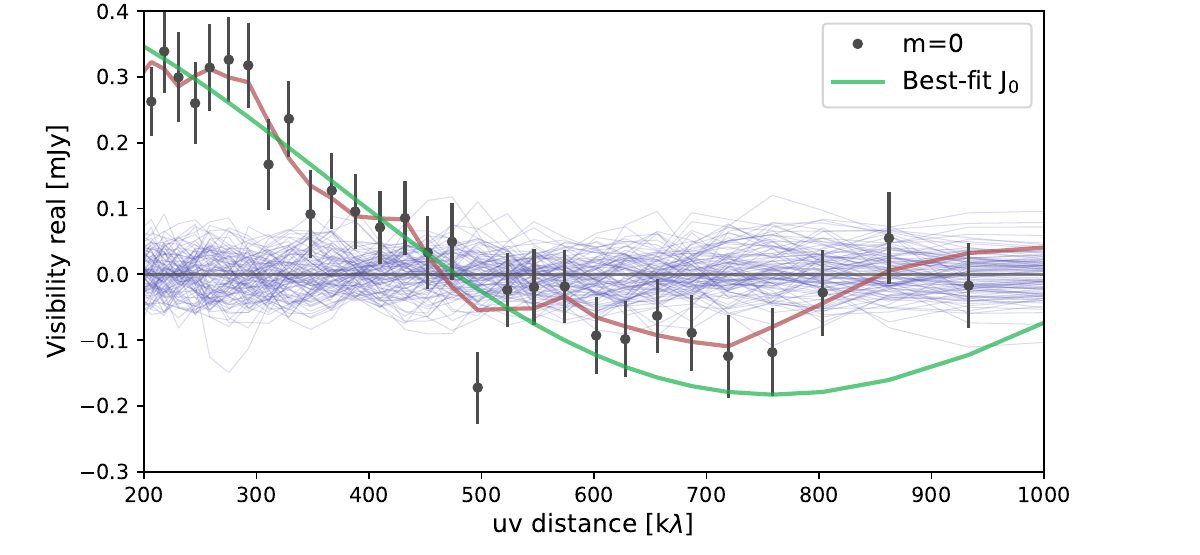} \\
    \includegraphics[width=10.5cm]{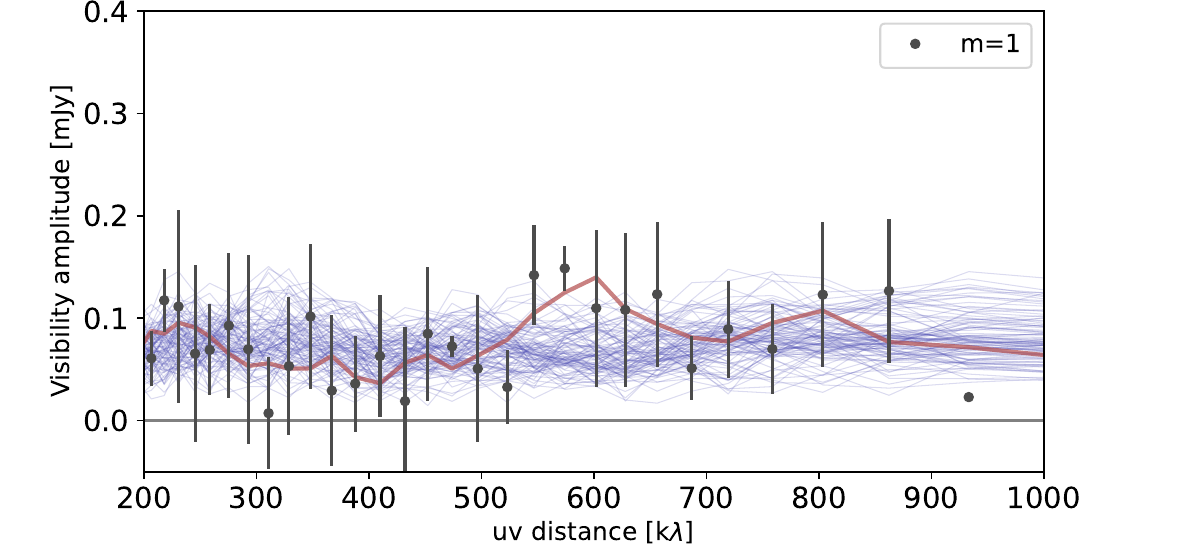} 
    \end{tabular}
    \caption{
    Visibility analyses for DM~Tau (Section \ref{sec:results}; Appendix \ref{appendix:asym}). 
    Data points in the upper and lower panels show the real part of the m$=$0 moment (monopole) of the complex visibilities and the amplitudes of the m$=$1 moment of the complex visibilities, respectively. Error bars were calculated by the standard deviation at each radial bin of the {\it uv}-space. 
    Red lines show the smoothed data points using the Savitzky-Golay filter (length of the filter window$=$7, order of the polynomial$=$2). Blue lines in each panel show 100 Savitzky-Golay filter smoothed random realization of noises (note that visibility amplitudes are positive definite).
    Green line in the upper panel shows the best-fit of the 0$^{\mbox{\scriptsize th}}$ order Bessel function of the first kind ($J_{0}$). 
    }
    \label{fig:dipole}
\end{figure}

Finally, to reference the locations of spatially resolved features in the JVLA observations, we utilized the high-angular-resolution ALMA 216--233~GHz images towards DM~Tau, which were detailed in \cite{Kudo2018ApJ...868L...5K} (c.f. \citealt{Hashimoto2021ApJ...911....5H}). 
Before comparing the ALMA and JVLA images, we corrected for the proper motions (R.A.: 11.788~mas~yr$^{-1}$ and Decl.: $-$18.361~mas~yr$^{-1}$) quoted from the Gaia Data Release 3 \citep{Gaia_2023}, using the \texttt{fixplanets} task of the Common Astronomy Software Applications (CASA; \citealt{McMullin2007ASPC..376..127M}) software package.

\subsection{JVLA data processing}\label{sec:reduc}

We manually calibrated the JVLA data following the standard strategy (see details below) using the CASA software package.
For all the JVLA observations we utilized, after implementing the antenna position corrections, weather information, gain-elevation curve, and opacity model, we bootstrapped delay fitting and passband calibrations, and then
performed a complex gain calibration. 
We applied the absolute flux reference to our complex gain solutions and then applied all the derived solution tables to the target source.

\begin{sidewaystable*}
\caption{Details of JVLA observations on DM~Tau.}\label{tab:obs}
\centering
{\footnotesize
\begin{tabular}{cccccccccccc}
\hline\hline             
Epoch & Time & Band & Array config. & On-source time & Beam size & RMS noise & Peak intensity & Flux density & Flux calib. & Gain calib. & BP calib. \\
 & (UTC) & Freq. (GHz) & {\it uv}-dis. (m) & (min) &  & ($\mu$Jy/beam) & (mJy/beam) & (mJy) &  & Flux (Jy) & Flux (Jy) \\
\hline
\hline
A-1\tablefootmark{a} & 2011 07 26         & C             & A         & 24  & 0\farcs56$\times$0\farcs34  & 8.0  & $\cdots$   & $\cdots$   & 3C138         & J0431$+$2037      & J0319$+$4130\\
    & 09:55:29--13:54:48    & 4.3--7.8  & 793--36584&       & (PA: -54$^\circ$)           &       &       &       &               & 2.36$\pm$0.003   & 20.6$\pm$0.06\\
A-2\tablefootmark{b} & 2019 09 28         & Q             & A         & 35    & 0\farcs054$\times$0\farcs044 & 23 & $\cdots$   & $\cdots$   & 3C138         & J0431+1731        & J0319+4130 \\
    & 12:30:50--13:40:54    & 40.0--47.8    & 692--34720&       & (PA: -3$^\circ$)            &       &       &       &               & 0.196$\pm$0.0007 & 19.9$\pm$0.076\\
A-3\tablefootmark{b} & 2019 10 06       & Q             & A         & 35    & 0\farcs046$\times$0\farcs041 & 24 & $\cdots$  & $\cdots$   & 3C138         & J0431+1731        & J0319+4130 \\
    & 11:02:14--12:12:14    & 40.0--47.8    & 1678--36475&       & (PA: -10$^\circ$)            &       &       &       &               & 0.186$\pm$0.0005 & 19.0$\pm$0.054\\
A-4\tablefootmark{b} &  2019 10 14      & Q             & A         & 35    & 0\farcs052$\times$0\farcs049 & 23 & $\cdots$   & $\cdots$   & 3C138         & J0431+1731        & J0319+4130 \\
    & 10:13:50--11:23:50    & 40.0--47.8    & 742--36617&       & (PA: 60$^\circ$)            &       &       &       &               & 0.191$\pm$0.0004 & 18.8$\pm$0.038\\    
\hline
B-1\tablefootmark{a} & 2011 02 25       & C             & B         & 3.5   & 1\farcs1$\times$1\farcs1  & 17  & $\cdots$   & $\cdots$  & 3C147         & J0449+1121        & 3C147\\
    & 01:52:16--03:51:46    & 4.1--7.9  & 121--11124&       & (PA: 20$^\circ$)            &       &       &       &               & 1.23$\pm$0.001   & ---\\
B-2\tablefootmark{a} & 2011 04 17--18   & C             & B         & 3.5   & 1\farcs1$\times$1\farcs0  & 25  & $\cdots$   & $\cdots$   & 3C147         & J0449+1121        & 3C147\\
    & 22:27:40--00:27:22    & 4.1--7.9  & 121--11124&       & (PA: 7$^\circ$)            &       &       &       &               & 1.02$\pm$0.002   & ---\\
B-3\tablefootmark{a} & 2011 05 14--15     & C             & BnA       & 3.5   & 1\farcs5$\times$0\farcs48  & 21  & $\cdots$   & $\cdots$   & 3C147         & J0449+1121        & 3C147\\
    & 23:11:10--01:10:48    & 4.1--7.9  & 121--22816&       & (PA: 62$^\circ$)            &       &       &       &               & 0.95$\pm$0.002   & ---\\
B-4 & 2012 08 12      & Q             & B         & 83  & 0\farcs28$\times$0\farcs17  & 26  & 0.19 & 0.56 & J0542$+$4951  & J0431$+$1731      & J0319$+$4130\\
    & 08:45:42--11:44:48    & 41.0--43.0    & 109--11124&       & (PA: -59$^\circ$)           &       &       &       &               & 0.224$\pm$0.0004 & 22.2$\pm$0.02\\
B-5 & 2012 08 29         & Q             & B         & 81  & 0\farcs28$\times$0\farcs16  & 24  & 0.18 & 0.57 & J0542$+$4951  & J0431$+$1731      & J0319$+$4130\\
    & 08:08:54--11:07:54    & 41.0--43.0    & 109--11124&       & (PA: -51$^\circ$)           &       &       &       &               & 0.228$\pm$0.0002 & 21.5$\pm$0.02\\
B-6 & 2012 09 04     & Q             & B         & 83  & 0\farcs22$\times$0\farcs17  & 26  & 0.17 & 0.59 & J0542$+$4951  & J0431$+$1731      & J0319$+$4130\\
    & 08:46:47--11:44:12    & 41.0--43.0    & 121--11124&       & (PA: -30$^\circ$)           &       &       &       &               & 0.223$\pm$0.0002 & 23.3$\pm$0.06\\
B-7 & 2012 09 15         & Q             & BnA       & 83  & 0\farcs20$\times$0\farcs12  & 25  & 0.14 & 0.70 & J0542$+$4951  & J0431$+$1731      & J0319$+$4130\\
    & 07:31:51--10:30:57    & 41.0--43.0    & 256--22816&       & (PA: -65$^\circ$)           &       &       &       &               & 0.249$\pm$0.0005 & 23.9$\pm$0.05\\
B-8\tablefootmark{a} & 2013 10 29         & X             & B         & 19  & 0\farcs97$\times$0\farcs82  & 5.9  & $\cdots$   & $\cdots$   & J0137$+$3309  & J0409$+$1217      & J0137$+$3309\\
    & 05:39:43--07:09:22    & 8.0--10.0 & 121--11124&       & (PA: -62$^\circ$)           &       &       &       &               & 0.343$\pm$0.0002 & $\cdots$ \\
\hline
C-1 & 2010 12 13       & Q             & C         & 126 & 0\farcs80$\times$0\farcs60  & 19  & 0.40 & 0.59 & J0542$+$4951  & J0431$+$1731      & J0319$+$4130\\
    & 01:14:16--05:13:21    & 41.0--43.0    & 43--3387  &       & (PA: -53$^\circ$)           &       &       &       &               & 0.361$\pm$0.0005 & 15.8$\pm$0.04\\
\hline
D-1\tablefootmark{a} & Aug. 24, 2010         & K             & D         & 9.5   & 4\farcs5$\times$3\farcs4  & 53  & $\cdots$  & $\cdots$  & J0542$+$4951  & J0431$+$2037      & J0319$+$4130\\
    & 08:30:25--12:59:41    & 20.0--22.0 & 23--1031  &       & (PA: -61$^\circ$)           &       &       &       &               & 0.754$\pm$0.0005 & 21.1$\pm$0.02\\
D-2 & 2010 08 30      & Q             & D         & 9.0   & 1\farcs9$\times$1\farcs7  & 134 & 0.81 & 0.86 & J0542$+$4951  & J0431$+$2037      & J0319$+$4130\\
    & 10:06:36--14:05:51    & 41.0--43.0 & 40--1031  &       & (PA: 83$^\circ$)            &       &       &       &               & 0.434$\pm$0.0008 & 17.4$\pm$0.02\\
D-3 & 2010 09 11        & K             & D         & 17  & 4\farcs2$\times$3\farcs5  & 11  & 0.14 & 0.13 & J0542$+$4951  & J0431$+$2037      & J0319$+$4130\\
    & 08:20:06--12:18:50    & 20.0--22.0 & 22--1031  &       & (PA: -51$^\circ$)           &       &       &       &               & 0.743$\pm$0.0003 & 21.2$\pm$0.01\\
D-4 & 2010 09 12         & Ka            & D         & 13  & 2\farcs9$\times$2\farcs1  & 34  & 0.44 & 0.53 & J0542$+$4951  & J0431$+$2037      & J0319$+$4130\\
    & 09:15:23--11:42:34    & 31.7--33.7 & 23--1031  &       & (PA: -3$^\circ$)           &       &       &       &               & 0.631$\pm$0.0020 & 23.0$\pm$0.04\\
\hline
\end{tabular}
\tablefoot{
\tablefoottext{a}{Target source was not detected.}
\tablefoottext{b}{Target source was detected at a high-enough S/N for constraining flux density only after combining the three epochs of Q band observations taken in 2019. }
}
}
\end{sidewaystable*}

\begin{table*}
{\footnotesize
\caption{Archival ALMA observations}              
\label{tab:ALMAarchive}      
\centering                                      
\begin{tabular}{c c c c c c c c}          
\hline\hline                        
Band & UTC date & Frequency & Angular resolution & MAS & $F_{\nu}$  & Project \\
 & (YYYY MM DD) & (GHz) & ($''$) & ($''$) & (mJy) &   \\
(1) & (2) & (3) & (4) & (5) & (6) & (7) \\
\hline                                   
3 & 2014 08 15 & 95.70--97.69 & 0.65 & 12 & 9.9$\pm$0.3 & 2013.1.00647.S (PI: Gorti, Uma) \\
  &            & 97.51--99.49 &      &    & 10.5$\pm$0.3 & \\
  &            & 109.01--110.90 &    &    & 14.6$\pm$0.5 & \\\hline
3 & 2016 11 05 & 94.51--96.49 & 0.66 & 6.8 & 10.5$\pm$0.4 & 2016.1.01042.S (PI: Chandler, Claire) \\
  &            & 96.41--98.39 &      &     & 11.2$\pm$0.4 & \\
  &            & 106.51--108.49 &    &     & 14.6$\pm$0.7 & \\
  &            & 108.46--110.33 &    &     & 15.1$\pm$0.7 & \\\hline
4 & 2016 07 25 & 144.02--146.01 & 0.45 & 5.0 & 29.7$\pm$1.5 & 2015.1.00296.S (PI: Semenov, Dmitry) \\
  &            & 145.92--145.98 &      &     & 27.8$\pm$1.8 & \\
  &            & 146.58--146.64 &      &     & 29.7$\pm$2.0 & \\
  &            & 156.96--157.01 &      &     & 34.5$\pm$2.3 & \\
  &            & 158.17--158.23 &      &     & 37.4$\pm$2.7 & \\
  &            & 158.95--159.00 &      &     & 36.5$\pm$2.3 & \\\hline
9 & 2018 08 16 & 658.15--659.09 & 0.30 & 3.1 & 366$\pm$55 & 2016.1.00565.S (PI: Schwarz, Kamber) \\
  &            & 660.67--661.60 &      &     & 371$\pm$56 & \\
  &            & 674.68--676.67 &      &     & 387$\pm$58 & \\
  &            & 674.68--676.67 &      &     & 384$\pm$58 & \\
\hline                                             
\end{tabular}
\tablebib{
(1) Frequency bands in ALMA convention. (2) Dates of the quoted observations. (3) Frequency coverage of the observations. (4) Angular resolutions quoted from the ALMA data archive. (5) Maximum recoverable angular scale (MAS) quoted from the ALMA data archive. (6) Flux densities measured by performing 2D Gaussian fittings. (7) Project codes and PIs.
}
}
\end{table*}

For the Q band (40--48 GHz) observations taken in 2019 (Table \ref{tab:obs}), we additionally used the observations on 3C138 to calibrate the cross-hand delay and absolute polarization position angle and we used the observations on J0319+4130 (3C84) to calibrate the polarization leakage (i.e., D-terms).
In spite of the fact that 3C84 was shown to be weakly polarized in the Q band,  previous studies have demonstrated that the D-term solutions derived from it can still help in suppressing polarization leakage. These solutions can also yield polarization images that are reasonably  consistent with those that were calibrated by 
the more ideal D-term calibrators for the JVLA Q band observations (e.g., 0713+4349; \citealt{Liu2016ApJ...821...41L,Ko2020ApJ...889..172K}).
We did not detect significant Stokes Q, U, and V intensities at Q band and we omit them from any further discussion. 

We performed multi-frequency synthesis (nterms$=$2) imaging (\citealt{Rau2011A&A...532A..71R}) using the CASA \texttt{tclean} task. 
For individual epochs of observations, the achieved synthesized beams and root-mean-square (RMS) noise in the Briggs Robust$=$2.0 weighted images are summarized in Table \ref{tab:obs}.
We omitted  the observations taken from epochs D-1 and D-2 (Table \ref{tab:obs}), owing to the compromised image quality and high RMS noise compared to the other observations taken at proximate time and frequency.

We jointly imaged all C band (4--8 GHz) observations taken in 2011, which yielded a 5.3~$\mu$Jy~beam$^{-1}$ RMS noise.
The target source, DM~Tau, was neither detected in this image nor in the individual epochs of C band observations.
DM~Tau was also not detected in the X band (8--10 GHz) observations taken in 2013 (Table \ref{tab:obs}; c.f., \citealt{Zapata2017ApJ...834..138Z}).
These results stand contrast to the significant X and Ku band detections in 2019 (\citealt{Terada2023ApJ...953..147T}), which can only be explained by variability.
We tried jointly imaging some Q band (40--48 GHz) observations using various combinations of observations and tapering.
When concatenating the Q band data, the A, (B \& BnA), and C array configuration observations were re-weighed (by factors of 4, 2, and 1, respectively) to yield a synthesized beam that is approximately Gaussian. 
We created the fiducial Q band image by limiting the {\it uv} distance range to $>$200 $k\lambda$. This helps achieve a synthesized beam that is small enough for diagnosing asymmetries.
The 40--48 GHz flux densities are dominated by spatially compact sources and thus there is no issue of missing short-spacing.
Using Briggs Robust$<2$ weighting (e.g., Robust$=$0, 1) to yield smaller synthesized beams significantly enhanced the thermal noise level, which is adverse for discerning localized sources. 
The profiles of the Q band images are discussed below in the relevant context.


\section{Results} \label{sec:results}

Figure~\ref{fig:image} shows a comparison of the Q band (40--48 GHz; $
\sim$7 mm) continuum image generated from combining all the observations at $>$200 k$\lambda$ {\it uv} distances (Q-fiducial image, hereafter) and the ALMA 225 GHz image made at the same synthesized beam (Section \ref{sec:data}). 
In Appendix \ref{appendix:image}, we provide the Q band images that were made from 
(1) combining all the Q band data we utilized (Q-all image, hereafter), 
(2) combining all observations taken in the A array configuration in 2019 (Q-A image, hereafter), 
(3) combining all observations taken in the A, BnA, and B configurations (Q-AB image, hereafter), 
and (4) combining only the observations taken in the B and BnA array observations in late 2012 (Q-B image, hereafter).
{ In all these images, we only significantly detected the 40-48 GHz emission in a small region that is presented in the right panel of Figure \ref{fig:image}.}

The most prominent feature in the ALMA image is a ring at $\sim$25 au radii that is only weakly asymmetric.
In the $\sim$0\farcs035 resolution 225 GHz image published by \citet{Hashimoto2021ApJ...911....5H},
only a very weak over-intensity (namely blob~A, with a $\sim$1.2 contrast ratio) in the west can be robustly identified; a weaker blob~B may be present in the south (contrast ratio $\sim$1.1) which remains to be confirmed by observations that have a better image fidelity.
The contrast ratios of blobs A and B were smeared to be lower when smoothing the ALMA image to the synthesized beam of our JVLA image.
In Figure \ref{fig:image}, although a northern and a southern arcs in the $\sim$25 au ring can be marginally seen in the ALMA image, they were in fact artifiically created by the well known effect of synthesized beam smearing (see Equation 9 of \citealt{Doi2021ApJ...912..164D}), which can be examined by smoothing the ALMA image.

In contrast, the JVLA Q band continuum images are dominated by an incomplete ring (Figure~\ref{fig:image}, right panel).
In the Q band, in both the Q-A image taken in 2019 and the Q-B image taken in 2012, the intensity peaks located at the south, which are offset from the major axis (P.A.$=$157.8$^{\circ}$; \citealt{Kudo2018ApJ...868L...5K}) of the 225 GHz ring.
The consistency between these JVLA images is hard to explain if the resolved features were spuriously created due to thermal noise or imaging artifacts.
With a better brightness temperature sensitivity, the Q-B image additionally detected an overintensity in the north.
The Q band observations also detected emission sources that are distributed over the $\sim$25 au ring, which appear knotty in the Q-A or Q-AB images.
In the Q-A image, some knots in the southeast and east have comparable or higher peak S/N than the most significant clump found in the Q-fiducial image and thus cannot be attributed to imaging artifacts.

In the Q-fiducial image (Figure \ref{fig:image}), several knots are detected at $\sim$3--6$\sigma$ level in east to south within the 25~au ring.  
They occupy about half of the ring and contribute $\sim300~\mu$Jy out of the free-free (Section \ref{sub:sed}) subtracted Q-band flux density of $\sim$500~$\mu$Jy.  
Assuming the remaining 200~$\mu$Jy is distributed smoothly in the other half of the ring (corresponding to $\sim6$ beam elements), we consider that the brightest knot in the south ($\sim$ 60 $\mu$Jy) may be at least 100\% brighter than the opposite side of the ring.  
This is a lower limit of the level of asymmetry in 40--48~GHz.
In contrast, the asymmetry in the higher angular resolution (0\farcs035) ALMA 225 GHz image is only $\sim$20\% brighter than the opposite side of the ring (\citealt{Hashimoto2021ApJ...911....5H}).
Moreover, the blob~A reported by \citet{Hashimoto2021ApJ...911....5H}, in fact, is not co-located with the knots resolved in the Q-fiducial image. Also, "blob
{\it  B"} is also spatially offset from the brightest knot in the Q-fiducial image. 

\begin{figure}
    \hspace{-0.35cm}
    \includegraphics[width=9.5cm]{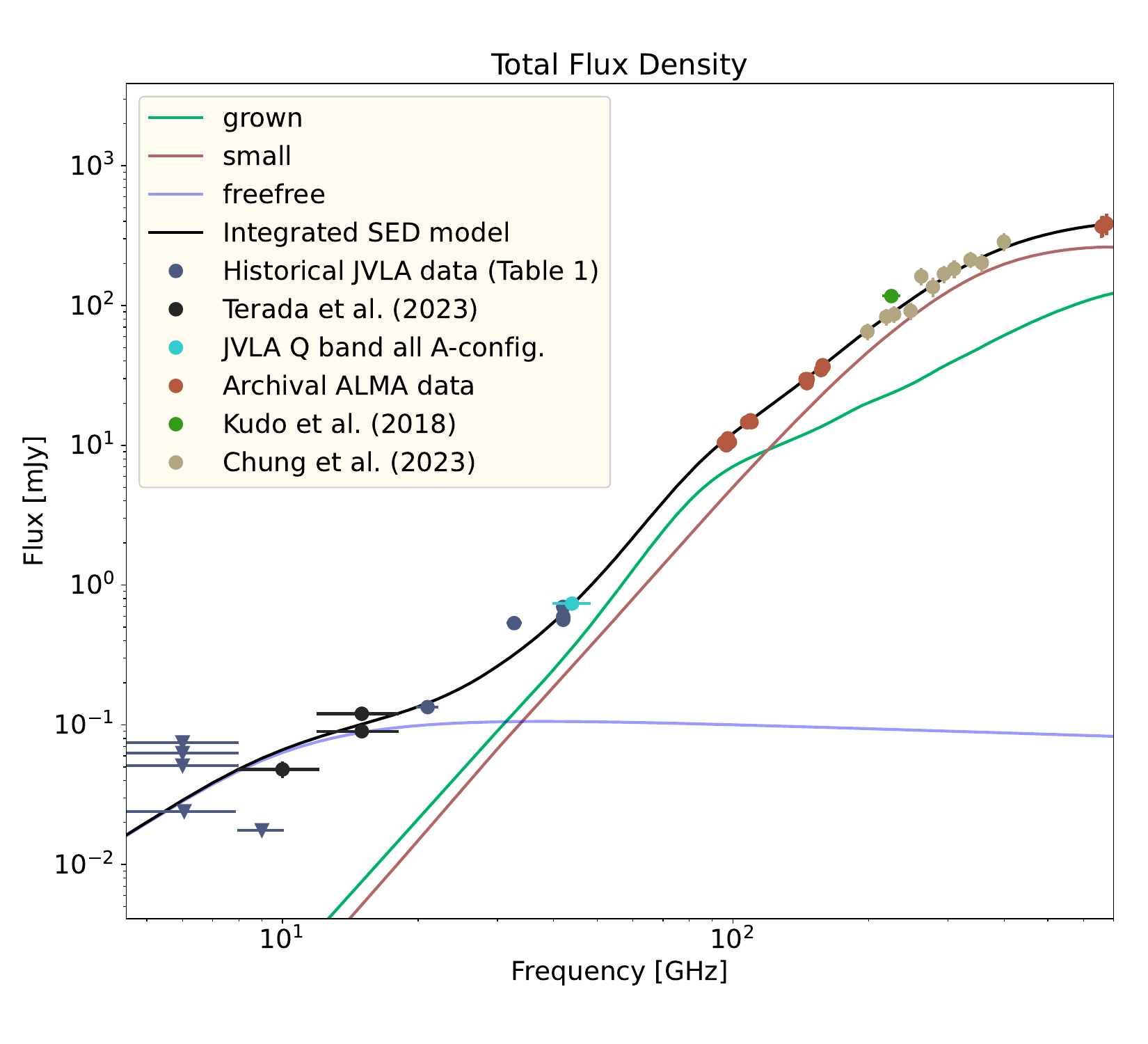}
    \caption{
    Flux densities of the DM~Tau disk ($\lambda=$0.43--67 mm). 
    Dots are the flux density measurements made from our own and the archival JVLA observations (for details see Section \ref{sec:data}), ALMA 225 GHz measurement from \citep{Kudo2018ApJ...868L...5K}, JVLA 8--12 GHz and 12--18 GHz measurements from \citep{Terada2023ApJ...953..147T}, archival ALMA Band 3, 4, and 9 observations (Table \ref{tab:ALMAarchive}), and the SMA 200--400 GHz measurements quoted from Chung et al. submitted.
    Triangles are the 3-$\sigma$ upper limits.
    The vertical error bars $\pm$1$\sigma$ error; the horizontal error bars show the frequency coverages of the measurements.
    { The sizes of some symbols are larger than the error bars.}
    Green, red, and blue lines show the flux densities of the grown dust, small dust, and free-free emission components in our best-fit model (Table \ref{table:model}; Section \ref{sec:discussion}) while the black line shows the integrated flux densities of these models.
    }
    \label{fig:sed_log}
\end{figure}

The Q-B image made with 41--43 GHz observations (Table \ref{tab:obs}) additionally revealed some emission slightly inward of the eastern part of the $\sim$25 au ring, which appears a lot dimmer in the Q-all image, which has a slightly smaller synthesized beam. 
The 41--43 GHz emission inward of the $\sim$25 au ring is likely to be tracing the time-varying free-free emission.
We cannot reliably analyze the 8--48 GHz spectral indices of the 40--48 GHz emission sources in the image domain due to the limited signal-to-noise ratio (S/N), and due to that the previous 8--12 GHz and 12--18 GHz images have indicated that both the flux density and position angles of the free-free emission sources may change day by day (\citealt{Terada2023ApJ...953..147T}).
We provide the quantitative estimates of the degree of free-free contamination in the 40--48 GHz images in Appendix \ref{appendix:freefree}.

The comparison of the Q-A and Q-B images indicates that there is a stationary, azimuthally asymmetric 40--48 GHz ($\sim$7 mm) emission component in the $\sim$25 au ring, whose origin is likely attributable to dust emission.
The presence of a (incomplete) ring can also be seen by directly analyzing the complex visibilities (for details, see Appendix \ref{appendix:asym}).
The upper panel of Figure \ref{fig:dipole} shows the real part of the $m=0$ moment (i.e., azimuthal averaging) of the complex visibilities taken from the Q band, B and BnA array configuration observations (Table \ref{tab:obs}).
Its shape is indeed consistent with that of the ring.
It crosses zero at $\sim$500 k$\lambda$ {\it uv}-distance, which corresponds to a $\sim$0\farcs15 ($\sim$23 au) ring radius. 
Fitting it with a 0$^{\mbox{\scriptsize th}}$ order Bessel function of the first kind (Appendix \ref{appendix:asym}) indicates that the radius of the ring is $\sim$0\farcs16 ($\sim$23 au) which agrees with the $\sim$25 au ring seen in the ALMA image.

We performed a model-independent assessment of the azimuthal asymmetry at Q band by examining the dipole ($m=1$) moments of the B and BnA configuration visibility amplitudes (Figure \ref{fig:dipole}) which are sensitive to the angular scales of the $\sim$25 au ring (for details see Appendix \ref{appendix:asym}).
Owing to the high noise, and due to the fact that the noise in the visibility amplitudes is positively biased, we are unable to robustly fit the dipole moments with a Bessel function. 
Nevertheless, from Figure \ref{fig:dipole}, it is still possible to visually identify weakly enhanced signals (with respect to pure noise) at 500--1000 k$\lambda$ with a peak at $\sim$600 k$\lambda$.
This indicates that the Q-Band intensity is azimuthally asymmetric on the spatial scale of $\sim 20$~au or less.
Higher sensitivity data will robustly confirm the presence of asymmetry.
The asymmetry may be caused by the non-uniform distribution of large grains within the $\sim$25~au ring or that of free-free emission sources (\citealt{Terada2023ApJ...953..147T}). 

Figure \ref{fig:sed_log} shows the 8--700 GHz SED of DM~Tau and some upper limits at lower frequencies.
We measured the 40--48 GHz flux density in 2019 from the Q-A image using aperture photometry.
The flux densities of the other observations were extracted using two-dimensional (2D) Gaussian fittings. 
The SED presents complicated frequency variations and thus cannot be appropriately described or discussed assuming a constant spectral index.
Instead, it signifies that there might be more than one emission sources, with one or some of them dominating the flux densities at certain frequency intervals.
A more detailed discussion about the nature of these emission sources is given in Section \ref{sec:discussion}.

\section{Discussion}\label{sec:discussion}

\subsection{Spectral index distribution and model}\label{sub:sed}

We are based on a spatially unresolved SED model to diagnose the properties of the observed dust emission sources.
The rationale of this approach is detailed in Appendix \ref{appendix:mcmc}.
In our models, the overall flux density is described by (quoting Equation 3 of \citealt{Liu2019ApJ...884...97L}):

\begin{equation}\label{eqn:multicomponent}
    F_{\nu} = \sum\limits_{i} F_{\nu}^{i} e^{-\sum\limits_{j}\tau^{i,j}_{\nu}}, 
\end{equation}
where $F_{\nu}^{i}$ is the flux density of the dust or free-free emission component, $i$, and $\tau^{i,j}_{\nu}$ is the optical depth of the emission component, $j,$ to obscure the emission component $i$ (e.g., $\tau^{i,j}_{\nu}=0$ if the component $i$ is not obscured by any other component in the line of sight).

For any emission component, the $F_{\nu}^{i}$ depends on the two free parameters,  temperature ($T$) and solid angle ($\Omega$).
For the free-free emission component, we adopted the approximated formulation outlined in \citet{Mezger1967ApJ...147..471M} and \citet{Keto2003ApJ...599.1196K}, where $F_{\nu}^{i}$ depends on another free parameter: the emission measure ($EM$).
We adopted the formulations and the default DSHARP size-dependent dust opacity tables published in \citet{Birnstiel2018ApJ...869L..45B} for the dust emission component.
The default DSHARP opacity table assumes that the dust grains are compact and composed of water ice \citep{Warren1984ApOpt..23.1206W}, astronomical silicates \citep{Drain2003ARA&A..41..241D}, troilite, and refractory organics (\citealt{Henning1996A&A...311..291H}.
The dust grain size ($a$) distribution $n(a)$ was assumed to be a power-law $\propto a^{-q}$ in between the minimum and maximum grain sizes ($a_{\mbox{\scriptsize min}}$, $a_{\mbox{\scriptsize max}}$), while it is 0 elsewhere. 
We assumed q=3.5 (\citealt{Mathis1977ApJ...217..425M,Doi2023arXiv230816574D}) and nominally fixed the value of $a_{\mbox{\scriptsize min}}$ to $10^{-4}$ mm, since the (sub)millimeter opacity depends very weakly on $a_{\mbox{\scriptsize min}}$.
In this case, the $F_{\nu}^{i}$  of dust emission components depend further on $a_{\mbox{\scriptsize max}}$ and dust column density ($\Sigma_{\mbox{\scriptsize dust}}$).

Given the large contrast between the morphology resolved in the ALMA 225 GHz and JVLA 40--48 GHz images (Figure \ref{fig:image}), it may be more appropriate to describe the SED with at least two dust components.
We found that the observed SED (Figure \ref{fig:sed_log}) can be approximated with a small dust component, a grown dust component, and a free-free emission component.
The free parameters of these components were optimized using the Markov chain Monte Carlo (MCMC) method, which are summarized in Table \ref{table:model}.
We regarded the parameters represented by the 50th percentile of the MCMC samples as the best-fit parameters.
The best-fit models are overplotted with the observational data in Figure \ref{fig:sed_log}, with the corner plots given in Figure \ref{fig:corner}.

From Figure \ref{fig:sed_log}, we see that the observed flux densities at $>$40 GHz are reasonably well explained.
However, the Ka band (31.7--33.7 GHz) data and a X band (8.0–10.0 GHz) non-detection are largely offset from the best-fit model, probably due to the variability of free-free emission.
At X band, this can be seen by comparing the non-detection in 2013 (\citealt{Zapata2017ApJ...834..138Z}) and the detection in 2019 (\citealt{Terada2023ApJ...953..147T}; Figure \ref{fig:sed_log}); the variability at Ka band has not been directly constrained since there was only one epoch of observations in 2010 (Table \ref{tab:obs}).
As a consequence of the poorly constrained variability, the free parameters of the free-free emission components were barely constrained (Figure \ref{fig:corner}). 

\begin{table*}
{\footnotesize
\caption{Model parameters}              
\label{table:model}      
\centering                                      
\begin{tabular}{ccccccc}          
\hline\hline                        
 &
$T$ &
$\Omega$\tablefootmark{a} &
$\Sigma_{\mbox{\scriptsize dust}}$ &  
$a_{\mbox{\tiny max}}$ &
EM &
$M_{\mbox{\scriptsize dust}}$
\\
 &
(K) &
($10^{-12}$ sr) & 
(g\,cm$^{-2}$) &
(mm) &
(cm$^{-6}$\,pc) &
$M_{\oplus}$
\\
 & (1) & (2) & (3) & (4) & (5) & (6) \\
\hline                                   
\multicolumn{7}{c}{Model with two dust components} \\
\multicolumn{7}{c}{small dust} \\
Initial\tablefootmark{b}  & 14.0 & 6.5 & 3.5 & 0.045 & $\cdots$ & 690 \\
MCMC\tablefootmark{c} & 12.4$^{+2.7}_{-1.7}$ &  8.5$^{+2.8}_{-2.6}$   &  2.51$^{+0.81}_{-0.56}$    &   0.055$^{+0.021}_{-0.028}$   & $\cdots$  & 640$^{+490}_{-300}$\\
\multicolumn{7}{c}{grown dust} \\
Initial\tablefootmark{b} & 14.0 & 2.5 & 10.0 & 0.35 & $\cdots$ & 750 \\
MCMC\tablefootmark{c} & 14.5$^{+6.1}_{-4.2}$ &  2.9$^{+1.3}_{-1.1}$  &  9.55$^{+3.9}_{-3.7}$   &   0.306$^{+0.195}_{-0.087}$   & $\cdots$  & 840$^{+870}_{-520}$ \\
\multicolumn{7}{c}{free-free emission} \\
Initial\tablefootmark{b} & 8000 & 4$\cdot$10$^{-3}$ & $\cdots$ & $\cdots$ & 5$\cdot$10$^{8}$ & $\cdots$ \\
MCMC\tablefootmark{c} & 8000 & 3.2$^{+2.2}_{-1.7}\cdot$10$^{-3}$ & $\cdots$ & $\cdots$ & 4.5$^{+3.5}_{-2.8}$$\cdot$10$^{8}$ & $\cdots$\\\hline
\multicolumn{7}{c}{Model with one dust component} \\                     
\multicolumn{7}{c}{dust} \\
Initial\tablefootmark{b}  & 14.0 & 6.5 & 3.5 & 0.045 & $\cdots$ & 690 \\
MCMC\tablefootmark{c} & 15.4$^{+0.4}_{-0.4}$ & 6.1$^{+0.3}_{-0.3}$ & 6.9$^{+0.1}_{-0.2}$ & 0.021$^{+0.02}_{-0.01}$ & $\cdots$ & 1280$^{+80}_{-100}$ \\
\multicolumn{7}{c}{free-free emission} \\
Initial\tablefootmark{b} & 8000 & 4$\cdot$10$^{-3}$ & $\cdots$ & $\cdots$ & 5$\cdot$10$^{8}$ & $\cdots$ \\
MCMC\tablefootmark{c} & 8000 &  2.9$^{+2.4}_{-1.2}\cdot$10$^{-3}$ & $\cdots$ & $\cdots$ &  4.0$^{+3.1}_{-2.0}\cdot$10$^{8}$ & $\cdots$ \\\hline
\end{tabular}
\tablefoot{
\tablefoottext{a}{1 sr $\sim$4.25$\times$10$^{10}$ square arcsecond.}
\tablefoottext{b}{Initial guesses of the fitting parameters.}
\tablefoottext{c}{Best-fit parameters.}
}
\tablebib{
(1) Dust temperature for dust emission components and electron temperature for free-free emission component. (2) Solid angle of the emission components. (3) Dust column density. (4) Maximum (dust) grain size. (5) Emission measure for free-free emission component. (6) Dust mass in units of Earth mass ($M_{\oplus}$).
}
}
\end{table*}



The solid angle of the small dust component is comparable to what can be expected from the ALMA 225 GHz image (Figure \ref{fig:image}, left panel).
The grown dust component has a smaller solid angle; this component may largely represent the azimuthally asymmetric, knotty 40--48 GHz intensity distribution resolved in the Q-A or Q-AB images (Figure \ref{fig:image}, \ref{fig:moreimage}).
The ALMA 225 GHz image only presents a weak azimuthal asymmetry (Figure \ref{fig:image}), likely due to the high optical depths of both the grown and small dust components.
Conversely, the large contrast between the morphology resolved at 225 GHz and 40-48 GHz means that DM~Tau cannot be optically thin at 225 GHz.

We only obtained a loose constraint on the $a_{\mbox{\scriptsize max}}$ of the small dust component, although we can be certain that it is below 100 $\mu$m.
The $a_{\mbox{\scriptsize max}}$ value of the grown dust component has the highest probability of being around 0.2 mm, while the best-fit value is slightly higher due to degeneracy. 
We note that this degeneracy in the $a_{\mbox{\scriptsize max}}$ value may not be entirely artificial.
While we only used one grown dust component in the model, the degeneracy in $a_{\mbox{\scriptsize max}}$ may reflect the fact that there are multiple optically thick dust concentrations that have various $a_{\mbox{\scriptsize max}}$ values.
Presently, the best-fit $a_{\mbox{\scriptsize max}}$ value (0.3 mm) for the grown dust component should be regarded as a conservative lower limit.
Due to the confusion of the time-varying free-free emission,  we cannot presently rule out the possibility that there are dust concentrations that the $a_{\mbox{\scriptsize max}}$ are $\gtrsim$1 mm and can fractionally contribute to significant emission at $\sim$20--40 GHz.
In the integrated SED (Figure \ref{fig:sed_log}), the excess of dust emission at $<$50 GHz does not look particularly obvious.
Nevertheless, the $<$50 GHz excess is very significant if we examine the SEDs of some specific locations, for example, one synthesized beam area that covers the location of peak in the Q-fiducial image (Figure \ref{fig:image}).

In our best-fit, the grown dust component ($a_{\mbox{\scriptsize max}}\sim$0.3 mm) may have a comparably larger dust mass than the small dust component ($a_{\mbox{\scriptsize max}}\sim$50 $\mu$m), although the mass in the grown dust component may become smaller once its $a_{\mbox{\scriptsize max}}$ is updated to larger values in the future observational studies.
The overall dust mass in our best-fit model is 1440$^{+1360}_{-820}$ Earth masses ($M_{\oplus}$). 
Large dust mass indicates a low gas-to-dust ratio for the disk to be stable against gravitational instability.
Our dust mass estimates may be regarded upper limits, since the default DSHARP dust are not as opaque as those that included graphitic carbons (e.g., \citealt{Ricci2010A&A...512A..15R,Woitke2016A&A...586A.103W}).
In addition, it is possible to reduce the dust mass in the small dust component if we model it as two separate components that have distinct $a_{\mbox{\scriptsize max}}$ values. If we make the $a_{\mbox{\scriptsize max}}$ of one of them close to $\sim$100 $\mu$m, the anomalous reddening effect (\citealt{Liu2019ApJ...877L..22L}) allows us to fit the (close to 2.0) spectral index observed at 200--400 GHz (Chung et al. submitted) with a dust column density that is considerably lower than our present best-fit value for the small dust component (Table \ref{table:model}).
Nevertheless, the differences in the assumed opacities should not change the orders of magnitude of our dust mass estimates; as discussed above, the small dust component is unlikely to be optically thin everywhere.
The grown dust mass budget in DM~Tau may be sufficient for feeding the formation of massive planets that the metal masses are as high as several 10$^{2}$ $M_{\oplus}$ (c.f., \citealt{Sato2005ApJ...633..465S,Thorngren2016ApJ...831...64T}).

Notwithstanding, we also tried fitting the observed SED with only one dust component, which is the approach we can take provided with only the unresolved SEDs, but without the spatially resolved image.
This fitting favors a small $a_{\mbox{\scriptsize max}}$ value of $\sim$21 $\mu$m while it yields a similar overall dust mass as what was given by the two-component fitting (Table \ref{table:model}).
This result of single-component fitting may be read with a caution, as we briefly discuss in Appendix \ref{appendix:mcmc}.

\subsection{Physical implications}\label{sub:implication}


\subsubsection{Asymmetry in protoplanetary disks}\label{subsub:asymmetry}

Physically, two general ideas have been proposed to explain asymmetries in disk rings. 
In the first, the entire ring is asymmetric, and dust in the entire ring drifts towards a single azimuthal pressure bump. 
Such an azimuthal pressure bump may be realized by a global scale vortex (\citealt{Birnstiel2013A&A...550L...8B}). 
The resulting dust asymmetries tend to have substantial contrasts. 
Possible examples in observations include IRS~48 (\citealt{Ninke2013Sci...340.1199V,Ninke2015ApJ...810L...7V,Ohashi2020ApJ...900...81O}) and HD 142527 (\citealt{Casassus2013Natur.493..191C,Kataoka2016ApJ...831L..12K,Satoshi2018ApJ...864...81O}). 
In the second scenario, the bulk of the ring is symmetric and asymmetries only arise in small localized regions due to certain properties, for example: small vortices. 
There can be more than one azimuthal dust concentration and part of the ring may not be affected by any dust concentration (i.e., in the smooth “background”). 
Such small-scale asymmetries may arise from hydrodynamic instabilities, as proposed in \citet{Li2020ApJ...892L..19L} and \citet{Huang2020ApJ...893...89HVortex}. 
DM~Tau and LkCa~15 (\citealt{Isella2014ApJ...788..129I,Jin2019ApJ...881..108J,Facchini2020A&A...639A.121F}), may be examples of this type.
The classification of the recently reported very-low-mass disks CIDA~1 (\citealt{Pinilla2018CIDA1,Hashimoto2023arXiv230811837H}) is not yet clear since the asymmetry, despite being detected by the JVLA observations, was not resolved at a high spatial dynamic range. 

Phenomenologically, we may also classify sources based on the morphology at 225 GHz and 40--48 GHz.
The first kind appears largely azimuthally asymmetric at both 225 GHz and 40--48 GHz (e.g., IRS~48, MWC~758; \citealt{Ninke2015ApJ...810L...7V,Ohashi2020ApJ...900...81O,Dong2018ApJ...860..124DMWC758,Casassus2019MNRAS.483.3278C}).
In each of the two cases, IRS~48 and MWC~758, the 225--345 GHz peak and the 30--50 GHz peak are closely associated with each other.
The second kind appears azimuthally symmetric or only weakly asymmetric, at both frequencies (e.g., GM~Aur, HL~Tau; \citealt{Carrasco2016ApJ...821L..16C,Carrasco2019ApJ...883...71C,Macias2018ApJ...865...37M}).

The DM~Tau dusty disk, which appears weakly azimuthally asymmetric and smooth at $\gtrsim$200 GHz and then appears highly asymmetric at $\lesssim$50 GHz, may represent a link between these two kinds.
It may provide a hint that at least some symmetric and asymmetric disks potentially have a unified origin.
For example, they may be simply related by optical depths, otherwise by their progress on developing azimuthally asymmetric $a_{\mbox{\scriptsize max}}$ distributions.
{
In both cases, the observations at lower frequencies (i.e., longer wavelengths) may tend to find the disks asymmetric.
In the former case, the lower optical depths at low frequency (e.g., $\lesssim$50 GHz) allow for the localized (sub)structures to be resolved at higher contrasts.
Substructures that harbor pebble-sized (e.g., mm--cm) dust may also appear more prominent than the ambient dusty disk structures at $\lesssim$50 GHz frequencies, due to the higher emissivity of pebble-sized dust at such frequencies.
Conversely, at a specific observing frequency, the apparent morphology of a disk that harbors substructures may evolve from weakly asymmetric to significantly asymmetric  if the small dust grains that are aptly coupled with the (azimuthally symmetric) gaseous disk are depleted over time.
The depletion of small dust grains reduces the overall optical depths; it also removes the azimuthally symmetric components. 
}
At $>$200 GHz, { we may hypothesize that} the morphology of DM~Tau, CIDA~1, and LkCa~15 may become more like the classical asymmetric sources resolved in the previous $\sim$225 or $\sim$345 GHz observations (e.g., HD 142527, AB Aur, SR~21, SY~Cha, MWC~758; Oph~IRS~48; \citealt{Hashimoto2011ApJ...729L..17H,Fukagawa2013PASJ...65L..14F,Ninke2013Sci...340.1199V,Satoshi2018ApJ...864...81O,Casassus2013Natur.493..191C,Tang2017ApJ...840...32T,Dong2018ApJ...860..124DMWC758,Casassus2019MNRAS.483.3278C,Yang2023ApJ...948..110Y,Orihara2023PASJ...75..424O}) when some small dust is depleted (e.g., by converting to grown dust and/or by dust and gas mass dispersal).
AB~Aur is indeed dominated by optically thinner dust emission that has a 3.34$^{+0.18}_{-0.18}$ spectral index at 200--400 GHz (Chung et al. submitted).

\subsubsection{Dust growth in an initially smooth background disk}\label{subsub:growth}
Thanks to the bigger size of the DM~Tau disk, the present, spatially well-resolved case study provides a clearer physical picture.
From Figure \ref{fig:image}, it appears that the grown dust ($\gtrsim$300 $\mu$m) were detected at the inner edge of the 20--120 au outer DM~Tau dusty ring.
Physically, this may be because that the $\gtrsim$100 $\mu$m sized dust becomes dynamically decoupled from the gaseous disk and rapidly migrated inward, until it is trapped by a pressure bump (\citealt{Zsom2010A&A...513A..57Z,Birnstiel2016SSRv..205...41B}).
{ In DM~Tau, outside of the water-snow line, the presence of grown dust appears to be linked to the dust azimuthal asymmetries.}
It is not yet clear to us whether the dust azimuthal asymmetries facilitate further dust growth in the icy world or the asymmetry is a consequence of dust growth and dust migration. 
One concordant interpretation for the creation of the dust cavity (or gap), dust trapping in the inner edge of the $\sim$25 au ring and the azimuthally asymmetric distribution of grown dust may be planet-disk interaction (e.g., \citealt{Dong2018ApJ...866..110D, Li2020ApJ...892L..19L} and references therein).
Grain growth may be further promoted at the dust traps if the dust-to-gas mass ratios are locally enhanced, which may be an important mechanism to facilitate the formation of comets or ice giants in cases where the water-ice coated dust grains are not particularly sticky (\citealt{Kimura2015ApJ...812...67K,Musiolik2019ApJ...873...58M,Steinpilz2019ApJ...874...60S,Pillich2021A&A...652A.106P,Arakawa2023ApJ...951L..16A}).
Inefficient coagulation of icy dust grains may also explain why the dust mass in the 20--120 au ring in DM~Tau has been largely preserved instead of being consumed to the formation of planetesimals or planets, and why the $a_{\mbox{\scriptsize max}}$ value in the 20--120 au ring remains $<$100 $\mu$m.

\section{Conclusion}\label{sec:conc}

DM~Tau is a representative transitional disk whose weakly asymmetric dusty outer ring at 20--120 au radii has been resolved in the previous, high-angular-resolution ALMA 225 GHz continuum image.
We carried out the 0\farcs06 ($\sim$8.7 au) resolution JVLA 40--48 GHz observations towards DM~Tau and retrieved the archival JVLA observations on DM~Tau that were taken after 2010.
Intriguingly, the JVLA 40--48 GHz observations resolved an incomplete, knotty, ring-like structure at the radii ($\sim$25 au) that have a maximized 225 GHz intensity.
Based on the resolved morphologies and the modeling for the 8--700 GHz SED, we found that the $>$225 GHz flux densities are dominated by optically thick dust thermal emission, while the $<$50 GHz flux densities are contributed by the dust thermal emission and time-varying free-free emission.
In addition, our fiducial SED model indicates that the incomplete knotty ring resolved at 40--48 GHz is composed of some high column density dust concentrations that the maximum grain size $a_{\mbox{\scriptsize max}}$ is greater than 0.3 mm; in contrast, the rather symmetric 20--120 au ring resolved at 225 GHz traces a small dust population that the $a_{\mbox{\scriptsize max}}$ is only as large as $\sim$50 $\mu$m.
The grown and small dust occupy different areas in the ring without apparent mutual obscuration.

Assuming the default DSHARP dust opacity, the overall dust mass in our SED model is 1440$^{+1360}_{-820}$ $M_{\oplus}$, which is approximately evenly shared by the grown ($a_{\mbox{\scriptsize max}}\gtrsim$300 $\mu$m) and small ($a_{\mbox{\scriptsize max}}\lesssim$50 $\mu$m) dust.
In the DM~Tau disk, the overall dust mass budget is likely largely preserved and is still sufficient for feeding to planetesimal and planet formation.
Outside of the water-snow line in DM~Tau, dust growth may not be rapid or efficient.
Only a portion of the dust has been converted to "grown" dust ($a_{\mbox{\scriptsize max}}\gtrsim$0.3 mm), which may have migrated inward until it is trapped close to the inner edge of the DM~Tau outer ring.
Planet-disk interactions or shear instability may further lead to an azimuthally symmetric pressure distribution or the formation of vortices, resulting in the observed knotty and azimuthally asymmetric distribution of the grown dust component.
As the small dust in the DM~Tau is gradually depleted (e.g., due to grain growth or disk gas and dust mass dispersal), its morphology at 225 GHz (e.g., ALMA Band 6) may come to resemble well known asymmetric transitional disks such as AB~Aur or HD142527.

\begin{acknowledgements}
We thank the referee for the useful suggestions. 
The National Radio Astronomy Observatory is a facility of the National Science Foundation operated under cooperative agreement by Associated Universities, Inc.
This paper makes use of the following ALMA data: ADS/JAO.ALMA \#2013.1.00498.S, \#2013.1.00647.S, \#2015.1.00296.S, \#2016.1.00565.S., \#2016.1.01042.S, \#2017.1.01460.S, \#2018.1.01755.S.
ALMA is a partnership of ESO (representing its member states), NSF (USA) and NINS (Japan), together with NRC (Canada), MOST and ASIAA (Taiwan), and KASI (Republic of Korea), in cooperation with the Republic of Chile. The Joint ALMA Observatory is operated by ESO, AUI/NRAO and NAOJ.
The Submillimeter Array is a joint project between the Smithsonian Astrophysical Observatory and the Academia Sinica Institute of Astronomy and Astrophysics, and is funded by the Smithsonian Institution and the Academia Sinica (\citealt{Ho2004ApJ...616L...1H}).
H.B.L. is supported by the National Science and Technology Council (NSTC) of Taiwan (Grant Nos. 111-2112-M-110-022-MY3).
This work was supported by JSPS KAKENHI Grant Number 23K03463 and 18H05441.
T.M. is supported by Yamada Science Foundation Overseas Research Support Program.
Y.H. was supported by the Jet Propulsion Laboratory, California Institute of Technology, under a contract with the National Aeronautics and Space Administration (80NM0018D0004).
\end{acknowledgements}

%
%


\bibliographystyle{aa}
\bibliography{main}

\begin{appendix}

\section{Images for the JVLA data taken in various array configurations}\label{appendix:image}
\begin{figure*}
    \begin{tabular}{ p{8.5cm} p{8.5cm} }
    \includegraphics[height=10cm]{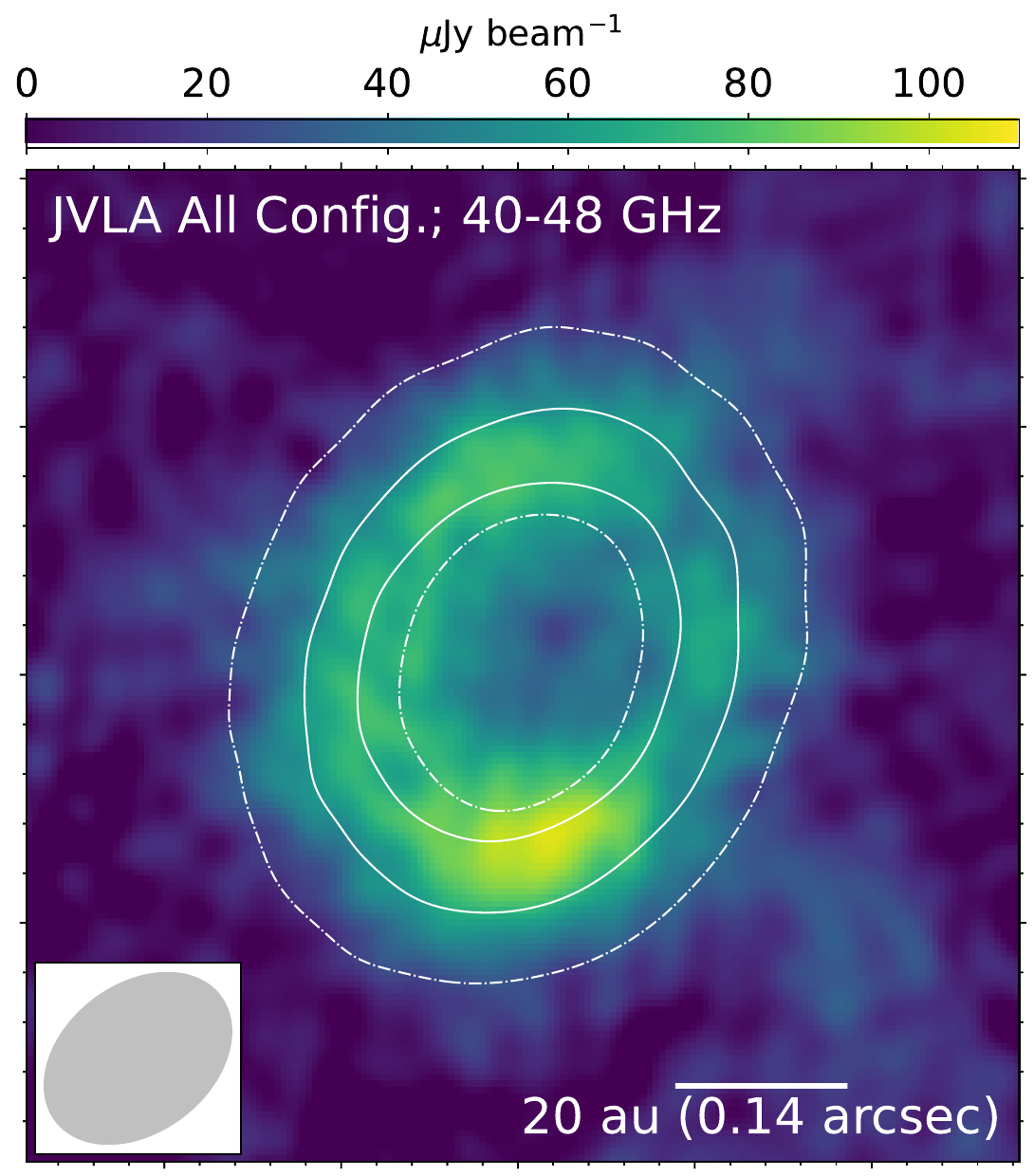} &
    \includegraphics[height=10cm]{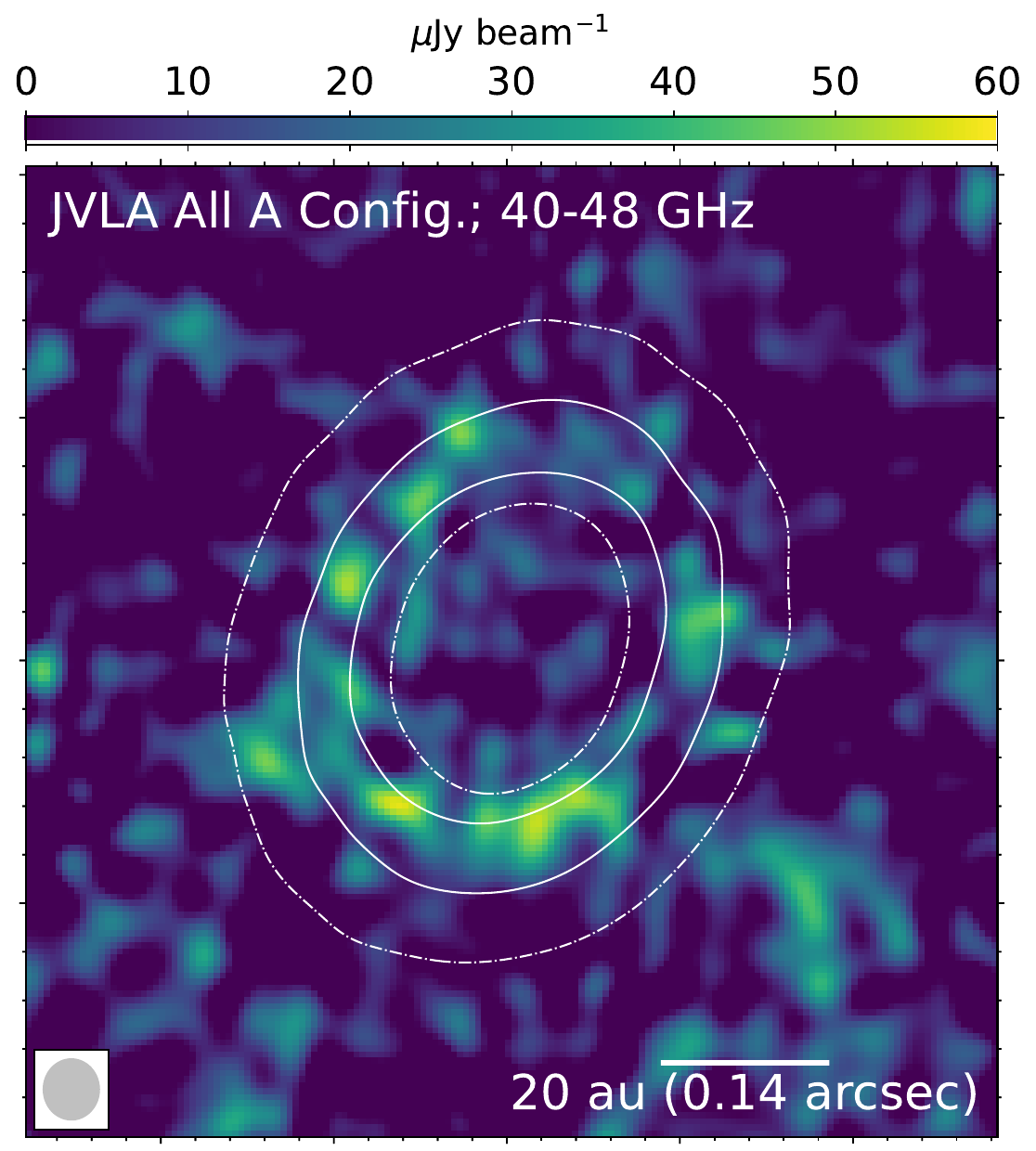} \\
    \includegraphics[height=10cm]{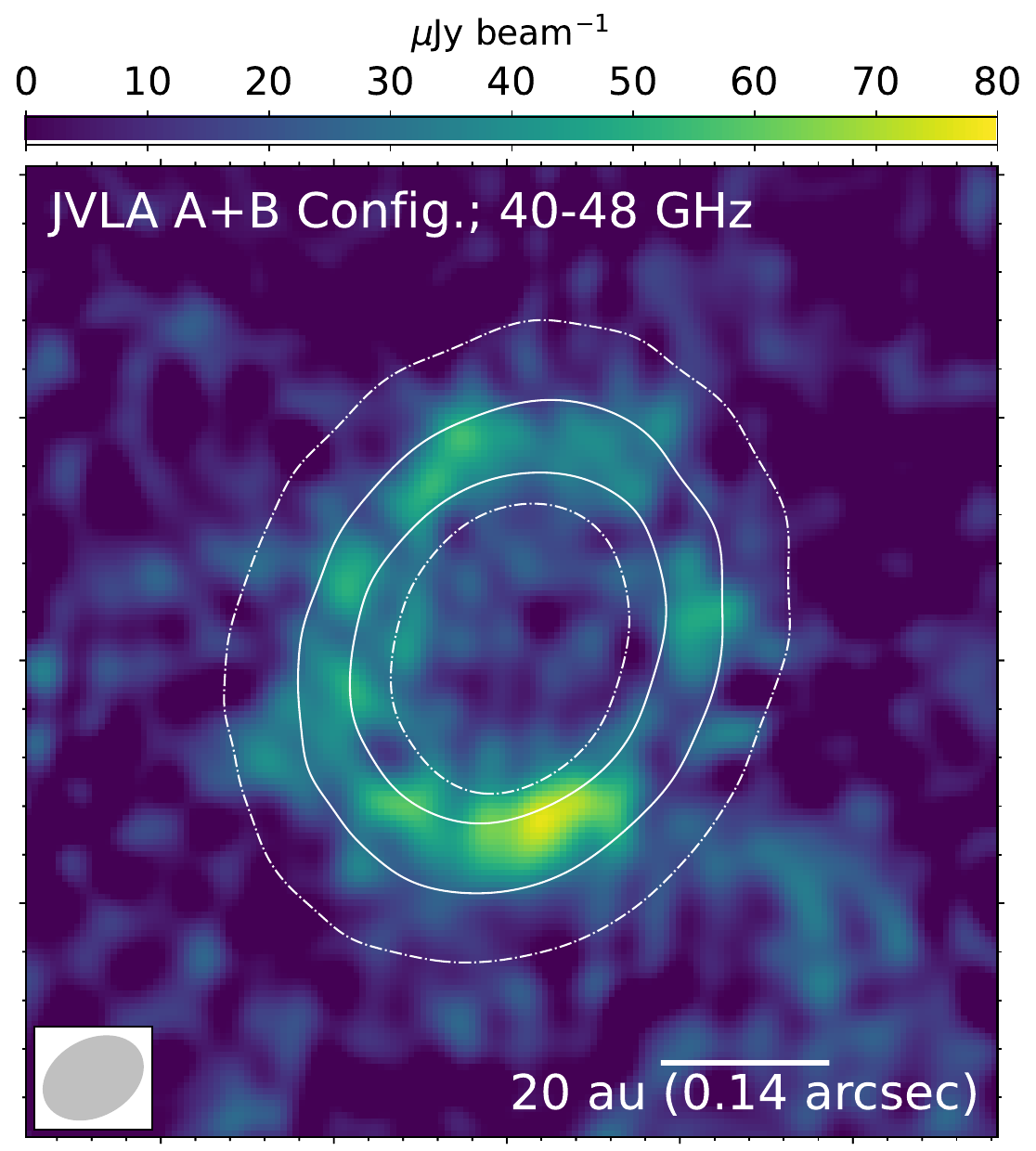} &
    \includegraphics[height=10cm]{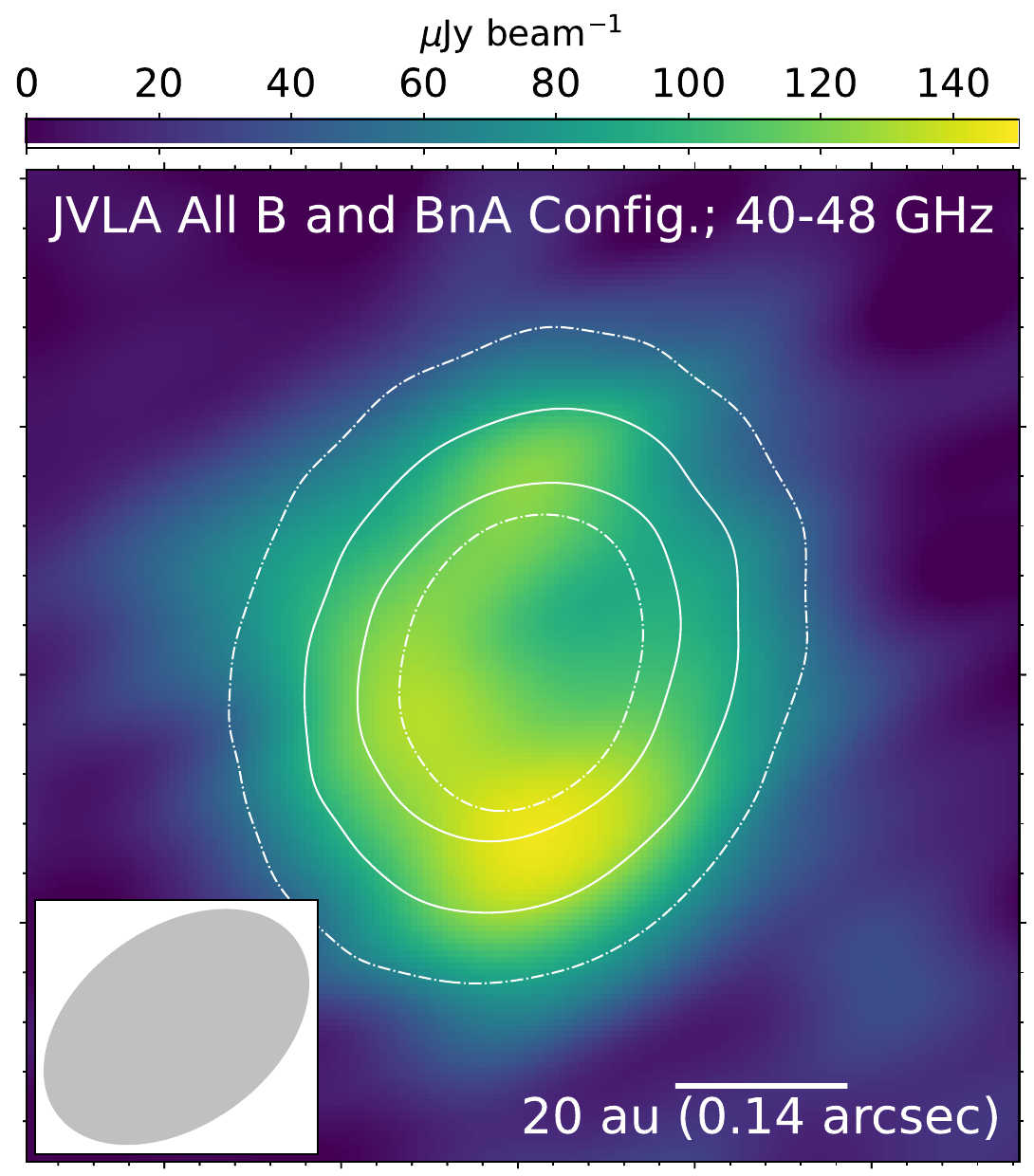} \\
    \end{tabular}
    \caption{
    Images on DM~Tau.
    {\it Top-left:}  JVLA Q band image produced by jointly imaging all Q band observations listed in Table \ref{tab:obs} (Q-all image; \beam=0\farcs17$\times$0\farcs12; P.A.=$-$53$^{\circ}$; RMS$=$8.1 $\mu$Jy\,beam$^{-1}$).
    {\it Top-right:}  JVLA Q band image produced by jointly imaging all Q band, A array configuration observations listed in Table \ref{tab:obs} (Q-A image; \beam=0\farcs074$\times$0\farcs058; P.A.=$-$63$^{\circ}$; RMS$=$9.7 $\mu$Jy\,beam$^{-1}$).
    {\it Bottom-left:} JVLA Q band image produced by jointly imaging all Q band, B, BnA, and A array configuration observations listed in Table \ref{tab:obs} (Q-AB image; \beam=0\farcs087$\times$0\farcs062; P.A.=$-$61$^{\circ}$; RMS$=$10 $\mu$Jy\,beam$^{-1}$).
    {\it Bottom-right:}  JVLA Q band image produced by jointly imaging all Q band, B and BnA array configuration observations listed in Table \ref{tab:obs} (Q-B image; \beam=0\farcs24$\times$0\farcs16; P.A.=$-$54$^{\circ}$; RMS$=$13 $\mu$Jy\,beam$^{-1}$).
    Contours in all panels are the same as those presented in Figure \ref{fig:image}.
    }
    \label{fig:moreimage}
\end{figure*}

Figure \ref{fig:moreimage} shows the JVLA 40--48 GHz images produced with various combinations of observational data. 

\section{Asymmetry}\label{appendix:asym}

Here, we briefly outline a model-independent method to detect asymmetric structures from visibility data.  A full description of the methods and the demonstration will be presented elsewhere as a separate paper.
The asymmetric structure in the image domain corresponds to the asymmetric distribution of the visibility in the {\it uv} domain; therefore, it is possible to assess the existence of the asymmetric structures by direct investigation of visibility.
If the real surface brightness distribution of the source on sky is $I(r,\theta)$, where $r$ is the distance from the central star and $\theta$ is the position angle measured from north to east, the visibility $V(\rho,\phi)$ is given by  
\begin{equation}
    V(\rho,\phi) = \int I(r,\theta) e^{ 2\pi i \rho r \sin(\theta+\phi)} r dr d\theta,
\end{equation}
where $\rho$ is the $uv$-distance and $\phi$ is the azimuthal angle in the $uv$-plane measured from the $u$-axis.

We define the $m^{\mbox{\scriptsize th}}$ multipole moment of the image as:
\begin{equation}
    \tilde{I}_m (r) = \frac{1}{2\pi} \int_{0}^{2\pi} I(r,\theta) e^{-im\theta} d\theta.
\end{equation}
If the surface brightness distribution is axisymmetric, $\tilde{I}_m = 0$ for $m\neq 0$.
The higher order $m$ modes correspond to the smaller scale asymmetry in the azimuthal direction.  
The brightness distribution on sky $I(r,\theta)$ is expressed as the Fourier series of $\sum \tilde{I}_m(r) e^{im\theta}$ over $-\infty<m<\infty$ and therefore the visibility distribution is given by:
\begin{equation}
    V(\rho,\phi) = \sum_{m=-\infty}^{\infty} e^{im\phi} 2\pi \int_{0}^{\infty} \tilde{I}^{\ast}_{m} (r) J_m (2\pi \rho r) r dr,
\end{equation}
where $J_m(x)$ denotes the $m^{\mbox{\scriptsize th}}$ order Bessel function and the asterisk denotes the complex conjugate.  
This indicates that the Hankel transformation of $\tilde{I}_m(r)$ gives the $m^{\mbox{\scriptsize th}}$ multipole moment of the visibility distribution of the $uv$-space.
If the 
$m^{\mbox{\scriptsize th}}$  moment of the asymmetric structure is located only at the radius $r_0$, $\tilde{I}_{m}(r)$ is proportional to $\delta(r-r_0)$.  In this case, the profile of the
$m^{\mbox{\scriptsize th}}$ moment of visibility is $J_m(2\pi \rho r_0)$.
Since the locations of zeros and peaks of Bessel function are well determined, it is possible to estimate the radial location of a ring (for 0-th moment) or a blob (for 1st or higher moments) using the locations of characteristic features (i.e., peak and zero) of the visibility profile, assuming that there is only one significant structure in radial direction.  
In Table~\ref{tab:BesselCharacter}, we summarize the relationship between the radial location $r_0$ on sky and the $uv$-distances of the first peak and the first zero $m=0$ and $1$.

In the analyses presented in Figure~\ref{fig:dipole}, 
we average the data over all the spectral windows in 40--43 GHz band and the integration over 30~s to improve the S/N and reduce the data size, while keeping sufficient number of data points in the visibility domain.
After deprojecting the data using $i=$35.2$^{\circ}$ and P.A.$=$157.8$^{\circ}$ \citep{Kudo2018ApJ...868L...5K}, the $uv$-plane is divided into 50 radial bins having similar data weights. 
The moments of the visibility are then calculated in each radial bin.

We note that incomplete $uv$-coverage, which is inevitable in interferometric observations, can introduce fake asymmetric structures and therefore the visibility data may need to be deconvolved for the $uv$-coverage.
For our dataset, we have checked that deconvolution does not affect our results and therefore, we present the analyses without deconvolution in this paper.
We will discuss more about how the incomplete $uv$-coverage affects the moment analyses in a separate paper that focuses more on analysis techniques. 

\begin{table}[h]
{\footnotesize
\caption{uv-distance scale of the first ($\rho>0$) zero-point and   the peak for $m=0$ and $m=1$ modes assuming that the structure in the sky plane is located at $r=r_0$.  The first peak of $m=0$ is at the same location with the first zero of $m=1,$ since $J_0(z) = -J^{\prime}_1(z)$.}              
\label{tab:BesselCharacter}      
\centering      
\begin{tabular}{ccc}
 &  $m=0$  & $m=1$  \\
 \hline
First zero [k$\lambda$] &
$789(r_0/0\farcs1)^{-1}$ &
$1258(r_0/0\farcs1)^{-1}$ \\
Second zero [k$\lambda$] &
$1258(r_0/0\farcs1)^{-1}$ &
$604(r_0/0\farcs1)^{-1}$ \\
\end{tabular}
}
\end{table}

\section{Free-free contamination in the 40--48 GHz images}\label{appendix:freefree}

\begin{figure*}
\hspace{0.25cm}
\begin{tabular}{ p{5.5cm} p{5.5cm} p{5.5cm} }
\includegraphics[width=6cm]{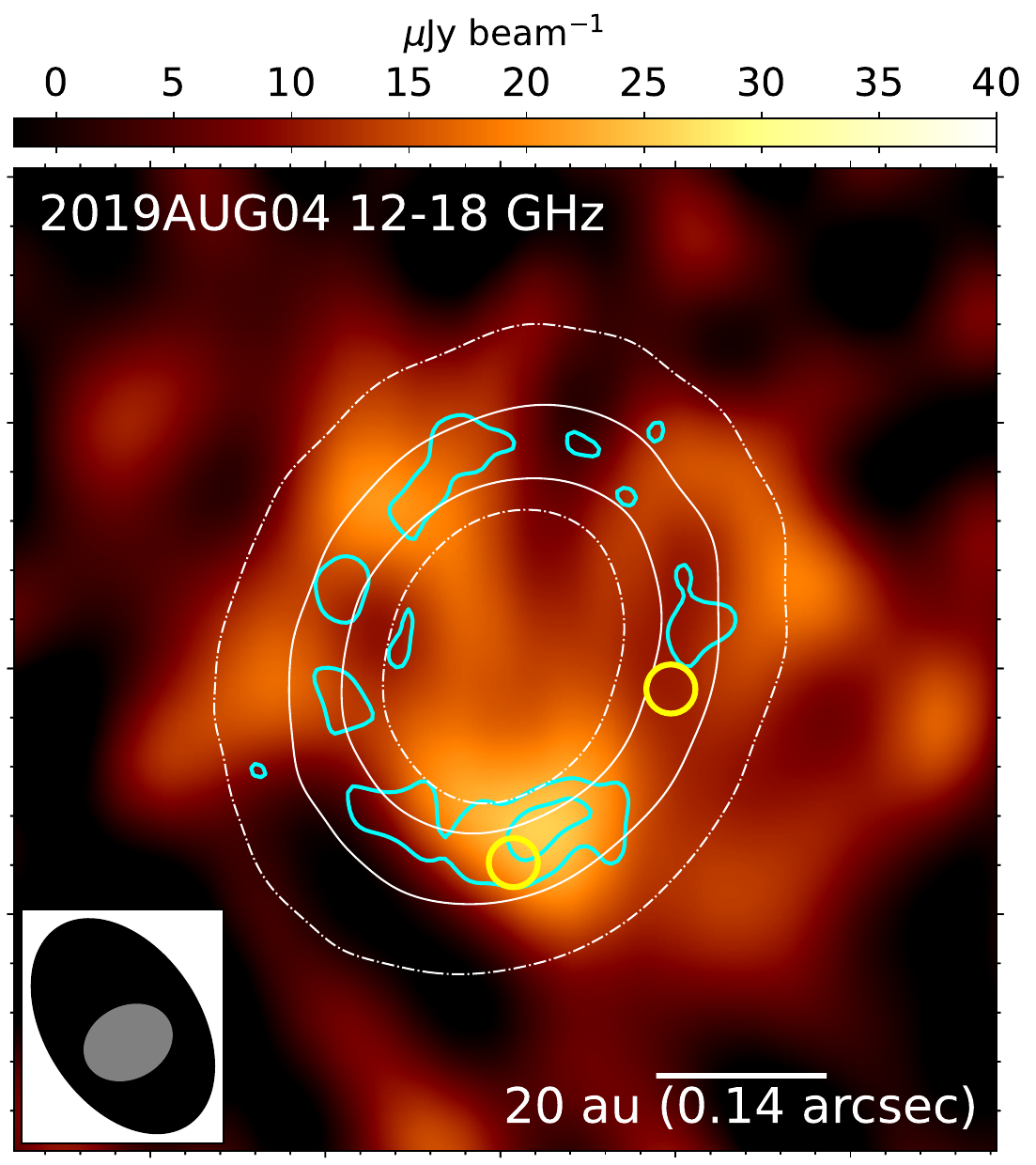} &
\includegraphics[width=6cm]{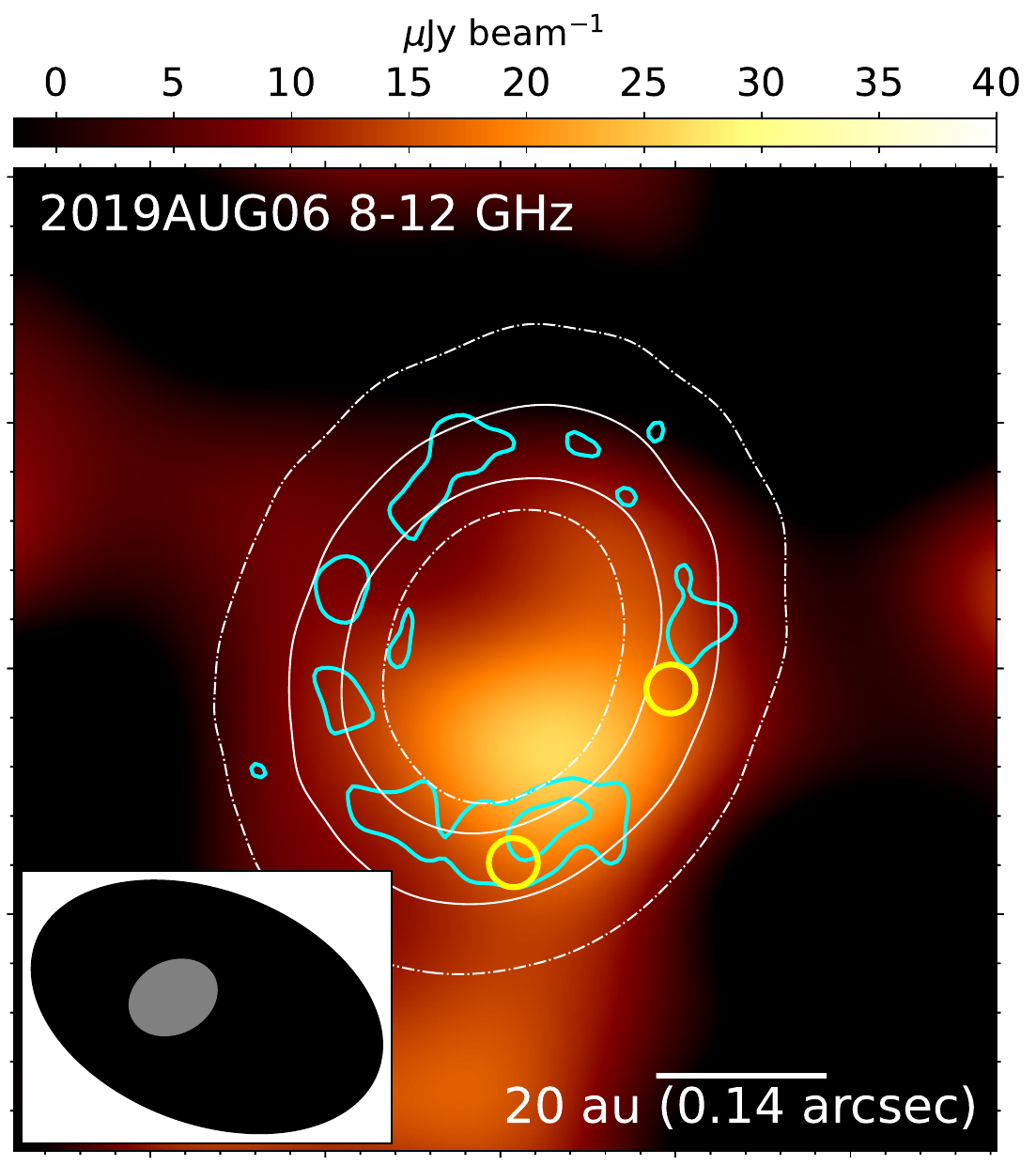}  &
\includegraphics[width=6cm]{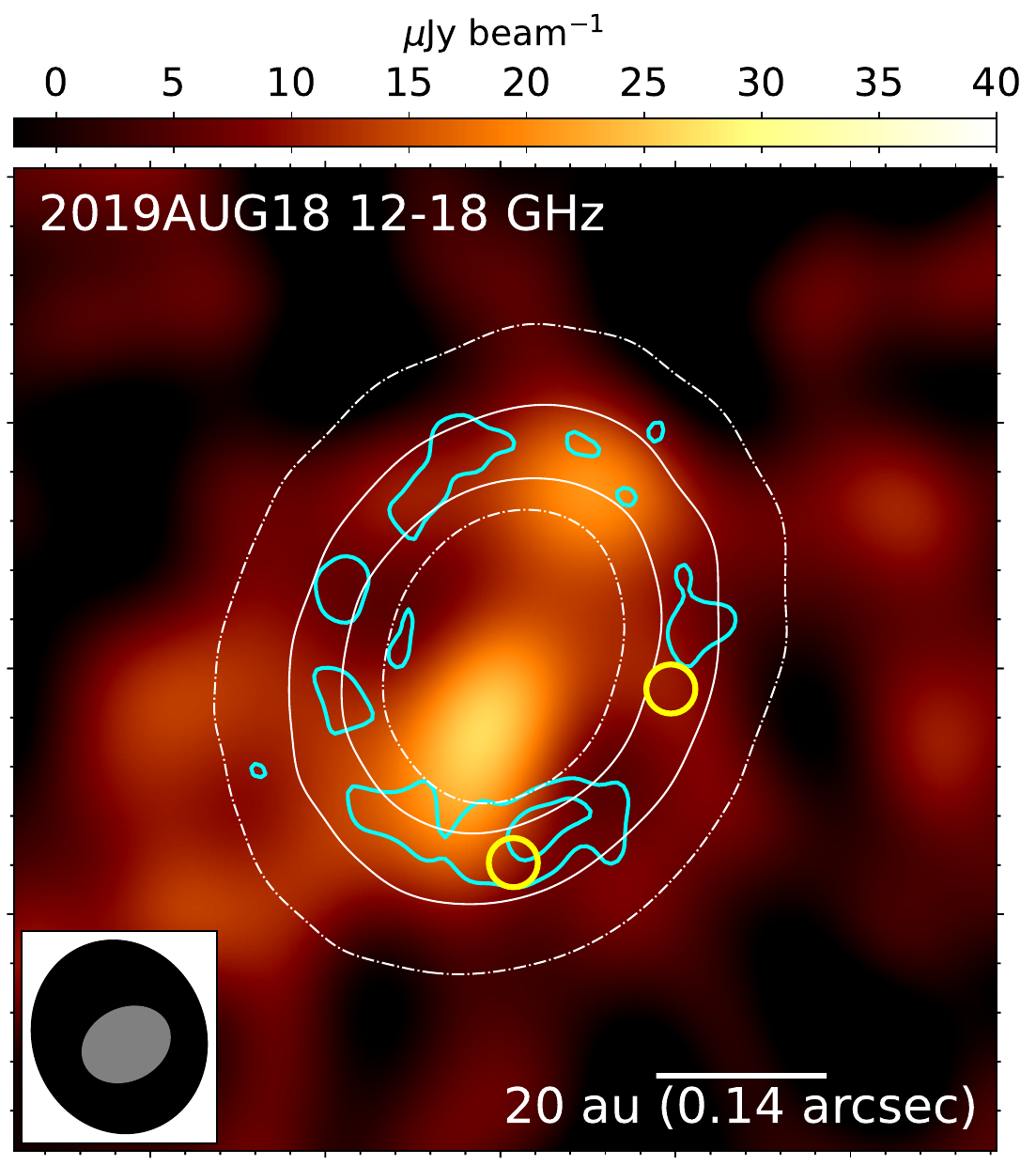} \\
\end{tabular}
\caption{
Comparisons of the ALMA image (white contours, same as the contours presented in the left panel of Figure \ref{fig:image}) and the JVLA images taken at various frequency bands.
Color images are the X band (8--12 GHz) and Ku band (12--18 GHz) images published in \citet{Terada2023ApJ...953..147T}; synthesized beams are shown in the lower left in black color.
Cyan contours shows the Q-fiducial image ($1\sigma\times$[3, 5]); synthesized beam is shown in the lower left in gray color.
Yellow circles marks the locations of the previously reported blobs~A and B (\citealt{Hashimoto2021ApJ...911....5H}).
}
\label{fig:freefreeimages}
\end{figure*}

In this section, we based on the morphology resolved in the Q-fiducial image and the intensity of free-free emission resolved in the previous X band (8--12 GHz) and Ku band (12--18 GHz) observations to assess the degree of free-free contamination in the Q-fiducial image. 

First, we point out that the knots resolved in the Q-fiducial image are spatially closely associated with the 25 au ring (Figures \ref{fig:image}).
In Figure \ref{fig:freefreeimages}, we compared the ALMA 225 GHz image and the Q-fiducial image with the X-band (8--12 GHz) and Ku-band (12--18 GHz) images reported by \citet{Terada2023ApJ...953..147T}.
Those low frequency observations mainly traced the time-varying free-free emission.
The morphology of the free-free emission changed day by day while the peak intensities, in terms of Jy\,beam$^{-1}$, were comparable ($\sim$30 $\mu$Jy\,beam$^{-1}$).
In two of the three epochs of the observations, the free-free emission peaks are located inside the 25 au ring.
If the free-free emission is strong enough to produce many knots resolved in our Q-fiducial or Q-A images, there should be a higher chance for us to detect 40--48 GHz emission knots at the area of the cavity instead of the 25 au ring. 
The close association of the 40--48 GHz emission knots (detected at $\gtrsim$3--5-$\sigma$) with the 25 au ring argues against the probability that these 40--48 GHz emission knots are mostly originate from free-free emission.
Free-free emission also cannot explain why the Q band, B-configuration and A-configuration observations that were widely separated in time (Table \ref{tab:obs}) consistently detected the brightest knot at the same location in the south.
Some weak Q band knots may or may not be confused by free-free emission knots.
If the other Q band knots were indeed confused by free-free emission, it would only strengthen the argument that the southern Q band knot was detected at a very high contrast and the 40--48 GHz intensity distribution is highly lopsided.

In addition, the free-free emission likely becomes optically thin at $>$10 GHz frequency (see Figure \ref{fig:sed_log}).
In this case, the spectral indices of free-free emission will be in the range of $-$0.1--0 at 10--48 GHz (e.g., \citealt{Anglada1998AJ....116.2953A}).
Therefore, in the Q band (40--48 GHz), the intensity of the brightest free-free emission knot will have a $\lesssim$30 $\mu$Jy\,beam$^{-1}$ peak intensity (Figure \ref{fig:freefreeimages}).
This is an upper limit of the peak intensity of one free-free emission knot, since the previous X and Ku band observations had coarse angular resolutions and thus might cover multiple free-free emission knots in one synthesized beam.
This means that there can be at most one free-free emission knot that can be detected at $\lesssim$3-$\sigma$ in our Q-fiducial image.
The other free-free emission sources should have lower intensities and thus cannot be detected in our Q-fiducial image. 
In other words, the $>$3-$\sigma$ sources in our Q-fiducial image more likely trace dust emission instead of free-free emission.

The estimates above do still overestimate the peak intensity of free-free emission in the Q-fiducial image.
In only one of the three epochs of  X- or Ku-band observations (i.e., the 12--18 GHz image taken on  2019 August 04, presented in the left panel of Figure \ref{fig:freefreeimages}), there was a free-free knot that is colocated with the Q band knot by chance.
This chance colocation alone is not sufficient to justify that the Q band knot is mainly contributed by free-free emission.
The intensity peak in the 12--18 GHz image taken on 2019 August 04 was only 26 $\mu$Jy\,beam$^{-1}$.
Supposing that in one of our three epochs of Q-band, A-configuration observations, the brightest Q band knot in the south would have also been confused with the free-free emission at a similar flux density.
After we combine the three epochs of Q-band, A-configuration observations, the flux density of the time-smeared free-free emission will become $\lesssim$9  $\mu$Jy\,beam$^{-1}$, which is lower than 1-$\sigma$ of our Q-fiducial image.
In the Q-fidicual and Q-A images, the knots detected at $\gtrsim$5-$\sigma$ have 50--70 $\mu$Jy\,beam$^{-1}$, which can hardly be explained by the much weaker free-free emission reported in \citet{Terada2023ApJ...953..147T}.

As a summary, we assess that locally, the Q band (40--48 GHz) intensity can be partly contributed by free-free emission.
However, the free-free emission is unlikely to be strong enough to produce many knots that can be detected in our Q-fiducial image.
Conversely, the $>$3-$\sigma$ detected 40--48 GHz knots in the Q-fiducial image that are closely associated with the brightest 225 GHz ring are most likely dust emission sources.
At Q band, the contribution of free-free emission can be prominent locally (i.e., in one synthesized beam of our Q-fiducial image) but should be a lot less significant as compared to dust emission in the integrated SED (Figure \ref{fig:sed_log}).

\section{SED modeling using MCMC}\label{appendix:mcmc}

Based on the ALMA 225 GHz observations and the complementary infrared data and assuming azimuthal asymmetry, \citet{Kudo2018ApJ...868L...5K} and \citet{Hashimoto2021ApJ...911....5H} have modeled the radial profiles of dust emission.
However, they did not take into consideration (sub)millimeter dust self-scattering, which has been known to be able to change the results considerably and usually makes the models more degenerate (\citealt{Liu2019ApJ...877L..22L}).
Incorporating (sub)millimeter dust self-scattering, with the frequency coverage, angular resolution, S/N, and image fidelity achieved by the existing observations on DM~Tau, our assessment is that a spatially resolved model would be too degenerated to provide any valid information.
Therefore, we followed the approach of \citet{Liu2019ApJ...884...97L} to model the integrated flux densities (Figure \ref{fig:sed_log}) as a spatially unresolved source, using components of dust and free-free emission.
Limited by the number of available independent measurements, we assumed that the physical properties within each component are uniform.
For the sake of robustness, we were restricted to use the smallest possible number of components that can approximate the observed flux densities. 
In this case, the model parameters for each component need to be regarded as the averaged properties of some emission sources that have similar SED shapes; thus, it is not necessary to focus on the face value of the number of the components we use.
For example, the knotty intensity distributions at 40--48 GHz (Figure \ref{fig:image}) signified multiple distinct dust concentrations, whereas we can only use one dust emission component to quantify their collective behavior. 
In addition, since we did not constrain the properties of the free-free emission using simultaneous observations at multiple frequencies, we could neither faithfully include more than one free-free emission components nor fit the individual epochs of $<$50 GHz observations by modeling with time-varying free-free emission (e.g., \citealt{Liu2021ApJ...923..270L}).

\subsection{Two-dust-component model (fiducial)}\label{appsub:twodust}

The simplest model that can be made without contradicting the distinct morphology resolved at 225 GHz and 40--48 GHz (Figure \ref{fig:image}) is a model based on two dust components.
Naturally, we would expect one of these two dust components to represent the bulk of the DM~Tau ring; the other presents the 40--48 GHz emission knots to some extent.
Similar to other recent case studies (e.g., \citealt{Liu2021ApJ...923..270L,Hashimoto2022ApJ...941...66H,Hashimoto2023arXiv230811837H}), we found that it is still possible to address some qualitative properties of dust emission based on such a simplified SED model (see discussion below).

\citet{Terada2023ApJ...953..147T} has resolved that the 8--18 GHz flux densities are dominated by the time-varying free-free emission.
In addition, the Q-all and Q-B images (Figure \ref{fig:moreimage}; Section \ref{sec:results}) show that not only dust but also free-free emission significantly contributed to the flux density at 40--48 GHz (Table \ref{tab:obs}) which also explains why the 8--18 GHz and 40--48 GHz flux densities varied (Figure \ref{fig:sed_log}).
Since the flux densities of the 8--18 GHz free-free emission had been well under 0.2 mJy, which are more than two orders of magnitude lower than the flux densities observed at $>$90 GHz frequencies.
Therefore, in our models, we required the flux densities observed at $>$90 GHz to be well-fit by stationary dust emission.
We required the $<$50 GHz flux densities to be roughly described by the emission of a free-free emission component and the stationary dust emission components. We realized the rough fitting at $<$50 GHz frequencies as follows:
\begin{itemize}
 \item[(1)] Setting the log likelihood to negative infinity when the model flux densities are (i) higher than the observed flux densities at 40--48 GHz in 2019, (ii) lower than the lowest observed 41--43 GHz flux density in 2012. This avoids the samplers that either do not produce significant dust emission or produce unrealistically high flux density at 40--48 GHz.
 \item[(2)] Omitting considering the C and X bands non-detections, and artificially setting the RMS errors (1-$\sigma$) of the X and Ku bands measurements to be identical to the observed flux densities (Table \ref{tab:obs}). The rationale is that the flux densities at these frequency bands are dominated by free-free emission that the variability (i.e., uncertainty) of flux densities appeared comparable to the flux densities (Figure \ref{fig:sed_log}). In our cases, the RMS noise of the observations were well below the uncertainties of flux densities.
\end{itemize}

Chung et al. (submitted) reported that the 200--400 GHz spectral index of DM~Tau is 2.11$^{+0.18}_{-0.17}$.
The spectral index becomes much lower at $>$400 GHz frequencies (Figure \ref{fig:sed_log}).
The directly measured spectral index between 400 GHz and 659 GHz was only $\sim$0.5.
To explain these low spectral indices, at $>$200 GHz frequencies, the dominant emission source is likely an optically very thick dust component that the averaged dust temperature is below the Rayleigh-Jeans limit.
In addition, it may require $a_{\mbox{\scriptsize max}}\lesssim$50 $\mu$m in this component, so that the anomalous reddening due to dust self-scattering can help suppress the spectral index at $\sim$700 GHz frequency (\citealt{Liu2019ApJ...877L..22L}).
We call this dust component the "small dust component" hereafter.
There is a weak degeneracy between the $a_{\mbox{\scriptsize max}}$ and $T_{\mbox{\scriptsize dust}}$ of the small dust component, which can be alleviated once the solid angle of the $\sim$700 GHz emission is better constrained by higher angular resolution observations.
Given that the dust temperature in the small dust component is constrained, and given that it is optically thick, its solid angle can be obtained by comparing the expected dust intensity (\citealt{Birnstiel2018ApJ...869L..45B}) with the observed flux densities (Figures \ref{fig:sed_log}).

\begin{figure*}
  \includegraphics[width=15.1cm]{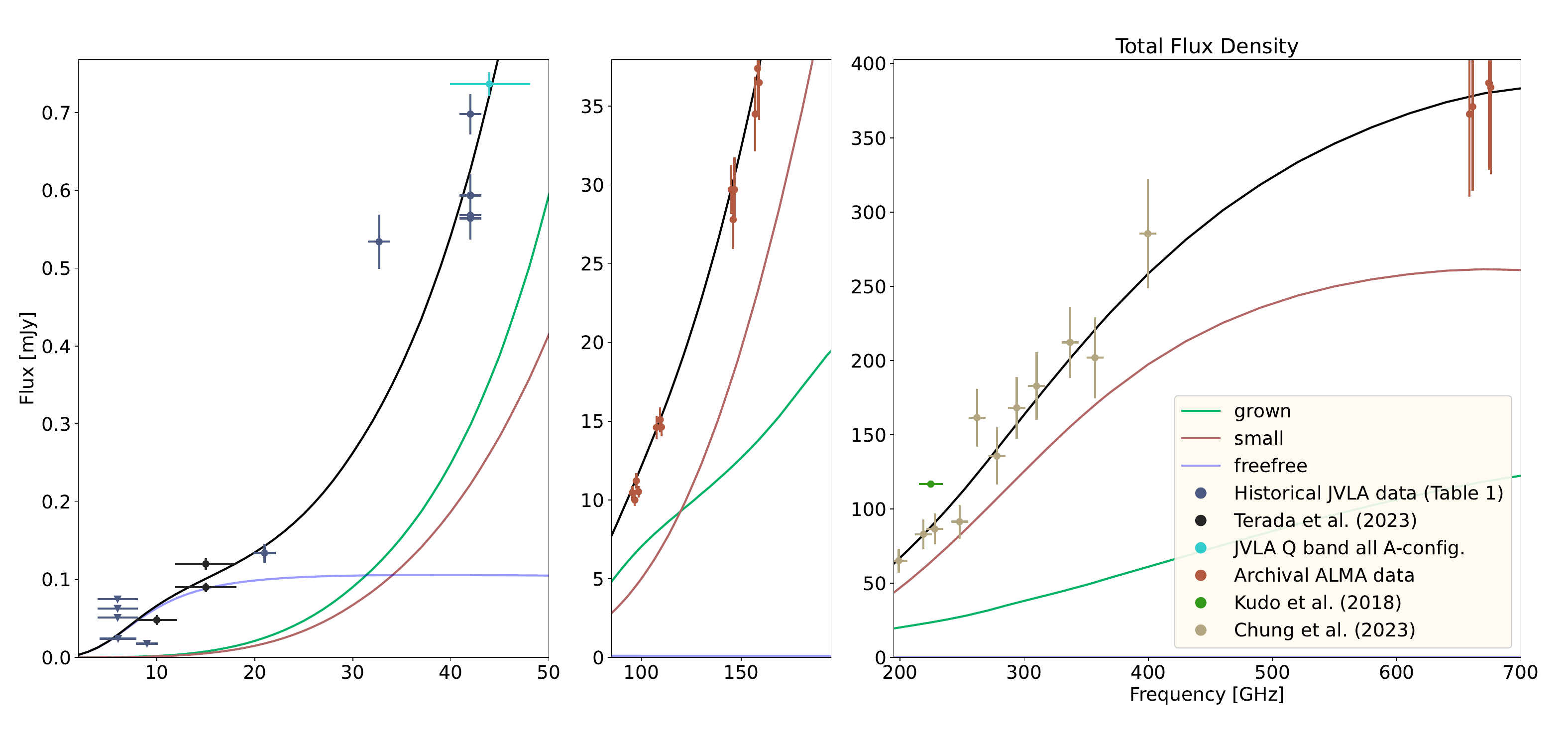}
  \vspace{-0.2cm}
  \caption{
  Similar to Figure \ref{fig:sed_log} but is presented in linear scale.
  }
  \label{fig:sed}
\end{figure*}

It turns out that the flux densities of DM~Tau at 95--110 GHz are too high and the spectral index between 110--160 GHz is too low to be consistent with the emission of the small dust component alone.
If we increase the $\Sigma_{\mbox{\scriptsize dust}}$ of the small dust component to make its flux density closer to what were observed at 95--160 GHz (Figure \ref{fig:sed_log}; Table \ref{table:model}), it will also contribute to high emission at 40--48 GHz, which will make tension with the JVLA Q band measurements. 
A better way to simultaneously fit the flux densities at 95--700 GHz, is to additionally include another optically thicker grown dust component that the $a_{\mbox{\scriptsize max}}$ is around 0.3 mm.
With $a_{\mbox{\scriptsize max}}\sim$0.3 mm, the SED of the grown dust component is anomalously reddened at $\sim$100 GHz due to dust self-scattering (c.f., \citealt{Liu2019ApJ...877L..22L}).
As expected, the MCMC fittings yielded distinct $a_{\mbox{\scriptsize max}}$ values for the two assumed dust components. 
In this case, the grown dust component can contribute to extra $\sim$100 GHz with a very low spectral index.
The results of MCMC fittings are presented in linear scale in Figure \ref{fig:sed}.

The co-existence of the small and grown dust components is consistent with the visual impression for Figure \ref{fig:image} (described in Sections \ref{sec:results} and \ref{sub:sed}).
Since the small dust component needs to have a high optical depth to explain the low spectral index observed at 200--400 GHz, we found that we cannot make the grown dust component obscured by the small dust component in the line-of-sight.
Otherwise, the attenuation will make it impossible for the grown dust component to contribute enough flux density at $\sim$95--160 GHz.

We optimized the free parameters in our model using the MCMC method, which was implemented with the \texttt{emcee} software package.
Without simultaneous, multi-frequency observations, the electron temperature ($T_{\mbox{\scriptsize e}}$) in the free-free emission component  cannot be constrained.
We nominally fixed $T_{\mbox{\scriptsize e}}$ to 8,000 K which is typical for ionized gas.
We found the initial guesses for the other free parameters interactively, which are summarized in Table \ref{table:model}.
Our prior assumptions for the free parameters are flat probability distributions in the linear domains.
These flat probability distributions are centered at our initial guesses and are upper and lower bounded by 10 and 0.1 times the initial guesses.
We used 250 samplers.
Before running MCMC, the samplers were randomly perturbed from our initial guesses based on the prior probability model. 
We advanced the positions of the samplers for 5$\times$10$^{5}$ iterations, discarding the first 10$^{4}$ iterations as the burn-in steps.

\subsection{Single dust component model}

\begin{figure*}
  \includegraphics[width=15.1cm]{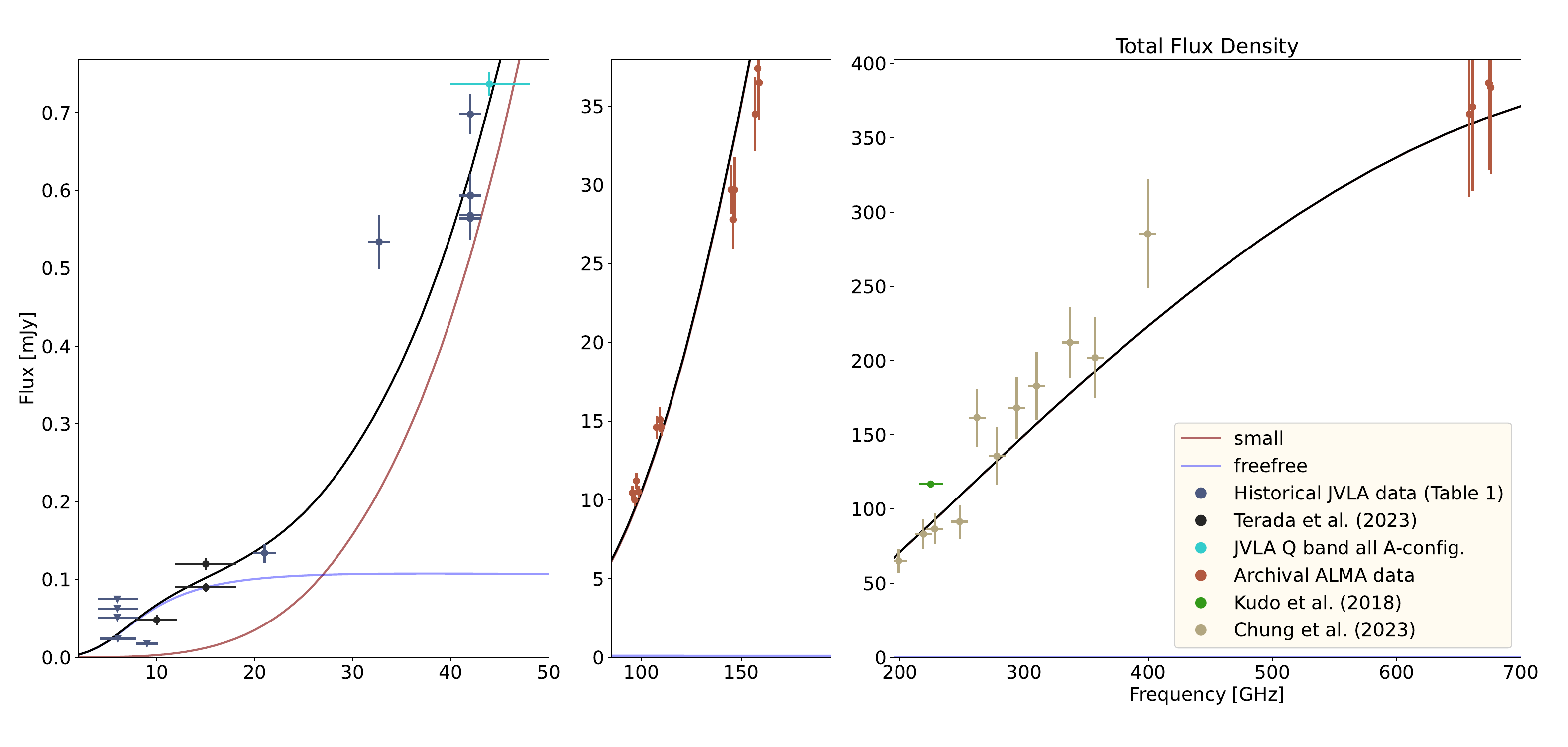}
  \vspace{-0.2cm}  
  \caption{
  Similar to Figure \ref{fig:sed} but is for the model with only a single dust component (see Appendix \ref{appendix:mcmc}).
  }
  \label{fig:sedSingle}
\end{figure*}

Using a similar approach as what was introduced in Appendix Section \ref{appsub:twodust}, we fit the observed SED assuming there is only a single dust emission component.
The single-component model (Figure \ref{fig:sedSingle}) cannot reproduce the 90--700 GHz data as closely as the two-component model introduced in the previous section, in spite that the marginalized posterior probability distributions appear confined (Figure \ref{fig:cornerSingle}).
It overestimated the flux densities at $\sim$150 GHz frequency and underestimated the flux densities at $\sim$100 GHz and $>$300 GHz frequencies.

In this case, due to the formal definition of the likelihood function (c.f., \citealt{Foreman-Mackey2013PASP}), the MCMC fittings favor a solution that is modestly deviated from every data point, instead of a solution that fit most data points excellently but is largely deviated from a few data points.
It is difficult to objectively determine which model is better.
This may be regarded as an ambiguity (or systematic bias) related to the choices in assessing the goodness-of-fit.
With another way of assessing the goodness-of-fit, the single-component model can potentially yield a very different results.

This ambiguity in the MCMC fitting can also be interpreted as our degree of freedom of choosing prior probability functions (\citealt{Foreman-Mackey2013PASP}).
For example, if we use a prior probability function to force the single dust component model to fit the $\sim$150 GHz data excellently, such a model will significantly underestimate the flux density at $\sim$100 GHz.
Physically, the means that there is some excess of low-frequency emission, which may be related to the presence of the knotty structures resolved at 40--48 GHz.


\begin{figure*}
    \hspace{0.3cm}
    \includegraphics[width=18cm]{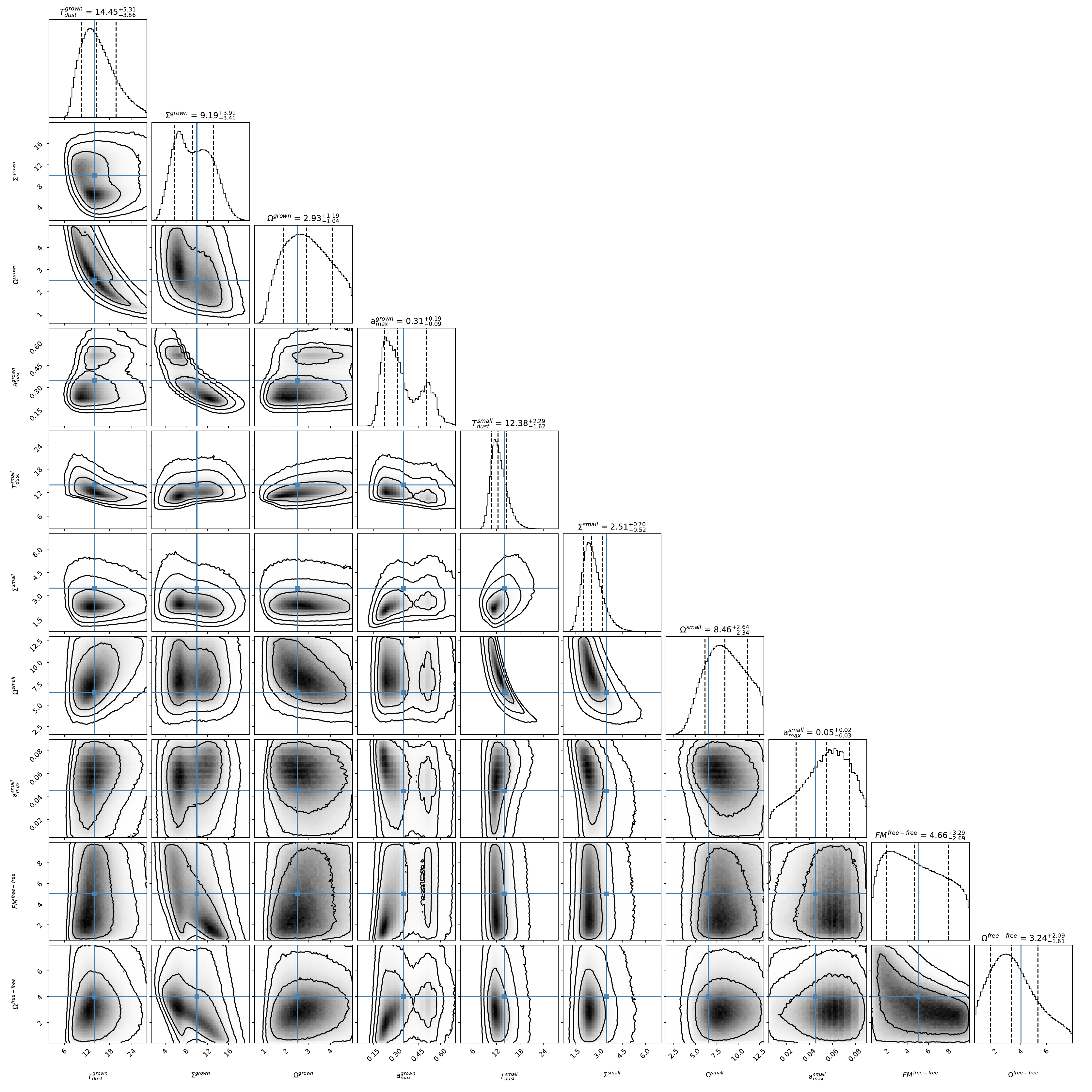}
    \caption{
    MCMC corner plot and the posterior distributions for the free-parameters in our SED model (Section \ref{sub:sed}; Appendix \ref{appendix:mcmc}; Table \ref{table:model}). 
    Units for all temperatures are Kelvin; units for dust column densities ($\Sigma$) are g\,cm$^{-2}$, units for $a_{\mbox{\scriptsize max}}$ are millimeter; units for $\Omega^{\mbox{\scriptsize grown}}$, $\Omega^{\mbox{\scriptsize small}}$, and $\Omega^{\mbox{\scriptsize free-free}}$ are 10$^{-12}$, 10$^{-12}$, and  10$^{-15}$ Sr, respectively; units for $EM$ is 10$^{8}$ cm$^{-6}$\,pc.
    The solid blue lines show the mean initial positions of the MCMC samplers. The 2D histograms show the projected distributions of the samplers over the planes defined by two free-parameters. The 1D histograms are marginal posterior distributions of these parameters. The vertical dashed lines indicated the 16th, 50th, and 84th percentiles of the samples, respectively, which are labeled on the top of each histogram. Some spikes in the 1D histograms are due to a few samplers that the positions cannot be advanced over iterations; however, they do not seriously impact our overall interpretation.
    }
    \label{fig:corner}
\end{figure*}

\begin{figure*}
    \hspace{0.3cm}
    \includegraphics[width=18cm]{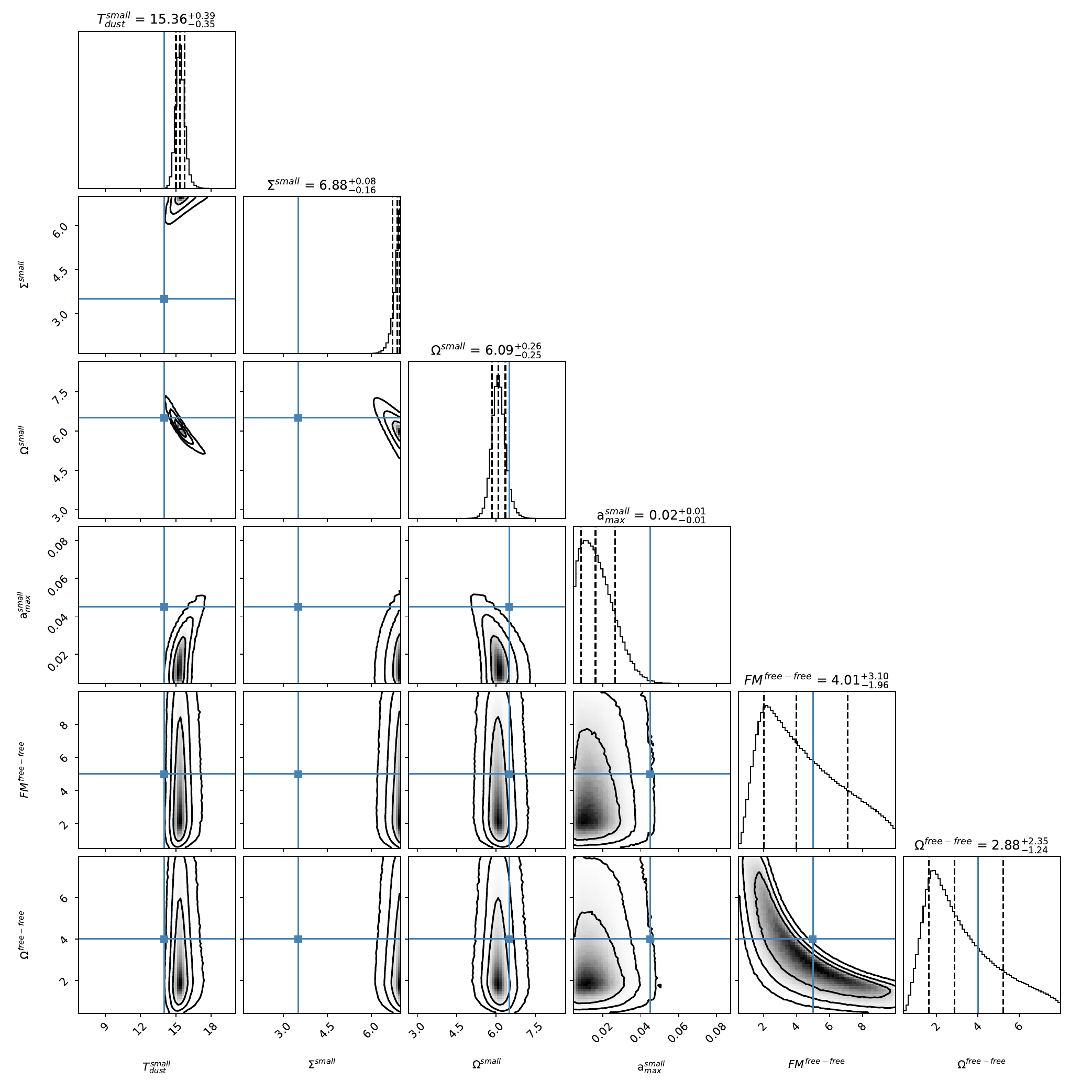}
    \caption{
    Similar to Figure \ref{fig:corner} but is for the model with only a single dust component (see Appendix \ref{appendix:mcmc}).
    }
    \label{fig:cornerSingle}
\end{figure*}

\end{appendix}

\end{document}